\documentclass[aps,
prb,
reprint,
floatfix,
nofootinbib,
]{revtex4-2}

\usepackage{graphicx}
\usepackage{physics}
\usepackage{booktabs}
\usepackage{amsthm,amssymb,mathrsfs}
\usepackage{tensor}
\usepackage{bm}
\usepackage[caption=false]{subfig}

\usepackage{feynmp-auto}
\usepackage{hyperref} 
\hypersetup{colorlinks=true,allcolors=blue}

\usepackage{xargs} 
\usepackage{xcolor}
\usepackage{slashed}
\usepackage{upgreek}

\usepackage[inline]{enumitem}

\let\temp\phi
\let\phi\varphi
\let\varphi\temp

\newcommand{\iu}{\mathrm{i}} 
\newcommand{\e}{\mathrm{e}} 

\newcommand{\DD}{\mathscr{D}} 
\newcommand{\D}{\mathrm{d}} 
\newcommand{\cc}{\mathrm{c.c.}} 
\providecommand{\ZZ}{\mathbb{Z}} 

\let\v\temp %
\let\temp\vv
\let\v\relax
\newcommand{\v}[1]{\ensuremath{\mathbf{#1}}} 

\makeatletter
\newsavebox{\@brx}
\newcommand{\llangle}[1][]{\savebox{\@brx}{\(\m@th{#1\langle}\)}%
	\mathopen{\copy\@brx\mkern2mu\kern-0.9\wd\@brx\usebox{\@brx}}}
\newcommand{\rrangle}[1][]{\savebox{\@brx}{\(\m@th{#1\rangle}\)}%
	\mathclose{\copy\@brx\mkern2mu\kern-0.9\wd\@brx\usebox{\@brx}}}
\makeatother

\makeatletter
\newcommand*{\coloneqq}{\mathrel{\rlap{%
			\raisebox{0.28ex}{$\m@th\cdot$}}%
		\raisebox{-0.28ex}{$\m@th\cdot$}}%
	=}
\newcommand*{\eqqcolon}{=\mathrel{\rlap{%
			\raisebox{0.28ex}{$\m@th\cdot$}}%
		\raisebox{-0.28ex}{$\m@th\cdot$}}%
}
\makeatother

\makeatletter
\newcommand*{\rom}[1]{\expandafter\@slowromancap\romannumeral #1@}
\makeatother

\usepackage{scalerel,stackengine}
\newcommand{\backvec}[1]{\reflectbox{$\vec{\reflectbox{\!$#1$}}$}}

\def\vecsign#1{\rule[1.388\LMex]{\dimexpr#1-2.5pt}{.36\LMpt}%
	\kern-6.0\LMpt\mathchar"017E}

\providecommand{\sfa}{\alpha}
\providecommand{\sfb}{\beta}
\providecommand{\sfc}{\gamma}
\providecommand{\sfd}{\delta}

\begin{document}
	
	\title{Fluctuation conductivity in ultraclean multicomponent superconductors}
	
	\author{Sondre Duna Lundemo} 
	\affiliation{Center for Quantum Spintronics, Department of Physics, Norwegian University of Science and Technology, NO-7491 Trondheim, Norway}
	
	\author{Asle Sudb\o}
	\email[Corresponding author: ]{asle.sudbo@ntnu.no}
	\affiliation{Center for Quantum Spintronics, Department of Physics, Norwegian University of Science and Technology, NO-7491 Trondheim, Norway}
	
	\date{\today} 
	
	\begin{abstract}
		We consider the intrinsic fluctuation conductivity in  metals with multiply sheeted Fermi surfaces approaching a superconducting critical point.
		Restricting our attention to extreme type-\rom{2} multicomponent superconductors motivates focusing on the ultraclean limit.
		Using functional-integral techniques, we derive the Gaussian fluctuation action from which we obtain the gauge-invariant electromagnetic linear response kernel.   
		This allows us to compute the optical conductivity tensor.
		We identify essential conditions required for a nonzero dissipative part of the longitudinal conductivity in a disorder-free and translationally invariant system.
		Specifically, this derives indirectly from the multicomponent character of the incipient superconducting order and the parent metallic state.
		Under these conditions, the enhancement of the DC conductivity due to fluctuations close to the critical point follows the same critical behavior as in the diffusive limit.  
	\end{abstract}
	
	\maketitle 
	
	\section{Introduction}\label{sec:intro}

    \begin{figure*}[t]
		\centering
		\includegraphics[width=0.65\textwidth]{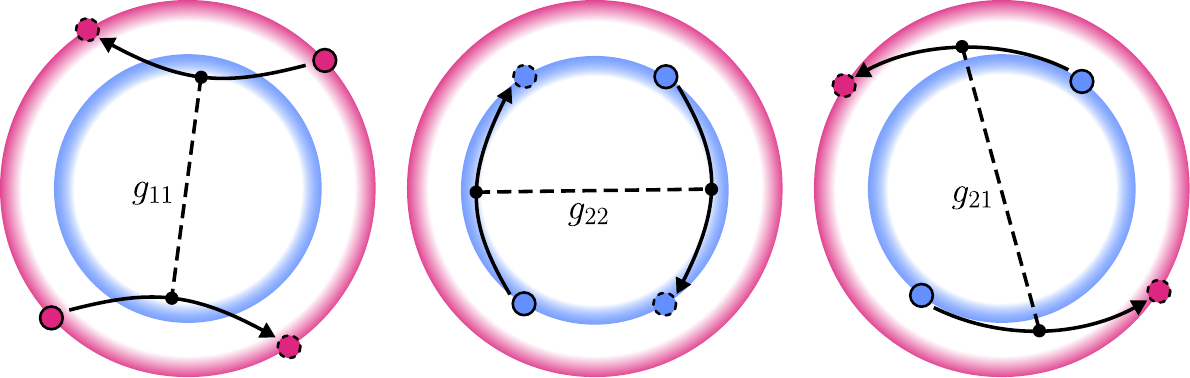}
		\caption{Illustration of the superconducting pairing allowed from the interaction considered in a two-band model. The blue and pink shells illustrate the Fermi surfaces of the two bands. The filled circles represent particles, and the dashed circles represent holes. 
		The dashed lines illustrate the pairing interaction $g_{\sfa\sfb}$.
		}
		\label{fig:multiband_scattering}
	\end{figure*}

    The collective behavior of interacting electrons in systems with multisheeted Fermi surfaces can result in novel physical phenomena.   
    In the presence of attractive effective interactions, electron pairing is the generic low-temperature instability of a Fermi liquid \cite{Cooper-Cooper-BoundElectronPairs-1956}.
    However, the resulting phenomenon of \textit{multicomponent} superconductivity has features with no counterpart in single-component superconductors, ranging from Leggett modes \cite{Leggett-Leggett-NumberPhaseFluctuationsTwoBand-1966} and spontaneous time-reversal symmetry breaking \cite{Stanev-Tesanovic-ThreebandSuperconductivityOrder-2010,Bojesen-Sudbo-TimeReversalSymmetry-2013,Lin-Hu-MasslessLeggettMode-2012,Takahashi-Hu-HTPhaseDiagram-2014,Yerin-vandenBrink-AnomalousDiamagneticResponse-2017} to exotic phases and phase transitions driven by competing topological defects \cite{Babaev-Ashcroft-SuperconductorSuperfluidPhase-2004,Smiseth-Sudbo-CriticalProperties$N$Color-2004,Smiseth-Sudbo-FieldTemperatureinducedTopological-2005,Herland-Sudbo-PhaseTransitionsThree-2010,Bojesen-Sudbo-PhaseTransitionsAnomalous-2014}.    
    The most intriguing physics of these systems takes place when fluctuations are included.
    Indeed, many of the exotic phases conjectured to appear in the vicinity of a multicomponent superconducting state -- for instance the sought-after charge-$4e$ superconductor -- require the inclusion of fluctuations \cite{Herland-Sudbo-PhaseTransitionsThree-2010,Bojesen-Sudbo-TimeReversalSymmetry-2013,Bojesen-Sudbo-PhaseTransitionsAnomalous-2014,Soldini-Neupert-Charge$4e$SuperconductivityHubbard-2024}.
    However, fluctuations may also leave observable signatures in electron transport properties, potentially offering insights into these elusive states.

    The study of fluctuation effects on superconductors above $T_c$ dates back all the way to the work of Ginzburg \cite{Ginzburg-Ginzburg-RemarksPhaseTransitions-1961}.
    The first accounts of their effects on transport coefficients were given in the pioneering works of Maki \cite{Maki-Maki-CriticalFluctuationOrder-1968}, Aslamazov and Larkin \cite{Aslamazov-Larkin-InfluenceFluctuationPairing-1968,Aslamazov-Larkin-EffectFluctuationsProperties-1968} and, later, Thompson \cite{Thompson-Thompson-MicrowaveFluxFlow-1970}, who showed how fluctuating Cooper pairs affect the electrical transport properties of the normal state. 
    With this, they brought together the fields of superconductivity and the theory of phase transitions.
    Since then, both thermal and electrical transport properties above $T_c$ have been studied extensively; see Ref.~\cite{Varlamov-Glatz-FluctuationSpectroscopyRayleighJeans-2018} and references therein.
    
    This topic later saw renewed interest driven by improvements in experimental capabilities \cite{Varlamov-Glatz-FluctuationSpectroscopyRayleighJeans-2018}. 
    The discovery of high-temperature superconductors, including the multicomponent iron-pnictide superconductors, also contributed in this respect, offering a much wider fluctuation region than before.
    That many of these materials were extreme type-\rom{2} superconductors motivated a reexamination of the fluctuation theory, which was originally derived for disordered superconductors \cite{Larkin-Varlamov-TheoryFluctuationsSuperconductors-2005}.
    Specifically, in these superconductors it is not unlikely that the electronic mean-free path considerably exceeds the correlation length of the fluctuation pairs, implying that the \textit{ultraclean} regime $ 1/(T\tau)^2 \ll \log(T/T_c) \ll 1 $ should be the most experimentally accessible \cite{Reggiani-Varlamov-FluctuationConductivityLayered-1991,Aronov-Larkin-GaugeInvarianceTransport-1995,Larkin-Varlamov-TheoryFluctuationsSuperconductors-2005}.
    Here, $\tau$ denotes the electron scattering time, $T$ is the temperature and $T_c$ is the mean-field critical temperature.
    However, the fluctuation theory of ultraclean systems proved surprisingly subtle \cite{Reggiani-Varlamov-FluctuationConductivityLayered-1991,Aronov-Larkin-GaugeInvarianceTransport-1995}. In particular, it was not clear whether the results could be obtained from the clean limit of an impure system or if the ultraclean limit should be treated separately from the outset. 
    In the present reexamination of the ultraclean fluctuation theory, we take the latter point of view and omit disorder altogether.

    Underlying all fluctuation calculations is always a set of approximations.
    A naive choice of approximation may conflict with the fundamental constraints on macroscopic observables set by conservation laws and symmetries in the system.
    In quantum many-body theory, these constraints manifest as Ward identities and are maintained by using approximations consistent with the two-particle irreducible effective action, also known as the Luttinger-Ward \cite{Luttinger-Ward-GroundStateEnergyManyFermion-1960} or Baym-Kadanoff functional \cite{Baym-Kadanoff-ConservationLawsCorrelation-1961,Baym-Baym-SelfConsistentApproximationsManyBody-1962}.
    Such \textit{conserving} approximations are indispensable for consistent calculations beyond perturbation theory. 
    In the context of electromagnetic response theory, gauge invariance poses a fundamental constraint on all physical quantities \cite{Schrieffer-Schrieffer-TheorySuperconductivity-1999,Rickayzen-Rickayzen-TheorySuperconductivity-1965}.
    A failure to respect gauge invariance may lead to spurious results, such as an unphysical Meissner effect in the normal state. 
    Indeed, as emphasized by Rickayzen \cite{Rickayzen-Rickayzen-TheorySuperconductivity-1965}, the current density is obtained as the small difference of two large terms, namely the paramagnetic and diamagnetic current densities. 
    Inconsistent approximations of these may obscure the cancellation of the leading terms and leave a finite diamagnetic response. 
    The issue of concern is, however, not peculiar to the study of superconductivity, rather it extends to all systems where interactions complicate the protection of local charge conservation \cite{Schrieffer-Schrieffer-TheorySuperconductivity-1999}.  
    
    That the established theory of ultraclean fluctuation conductivity was at odds with these fundamental principles was realized by Boyack in Ref.~\cite{Boyack-Boyack-RestoringGaugeInvariance-2018}.
    Only after including a new correction to the diamagnetic current density was the response kernel shown to respect gauge invariance and the spurious normal-state Meissner effect Rickayzen warned about was shown to disappear. 
    This diagram, together with those of the conventional fluctuation theory, were derived by using a functional-integral approach where fluctuations were included at the Gaussian level, thereby systematically reproducing the ladder approximation, i.e., the random-phase approximation for the fluctuation propagator \cite{Dupuis-Dupuis-FieldTheoryCondensed-2023}.
    Similar methods have also been employed to devise a consistent fluctuation theory below $T_c$, where the Meissner response is expected but collective modes in general have to be included to obtain a gauge-invariant response \cite{Wulin-Levin-ConductivityPseudogappedSuperconductors-2011,Guo-He-TheoriesLinearResponse-2013,Guo-Levin-FundamentalConstraintsLinear-2013,Anderson-Levin-GoingBCSLevel-2016,Boyack-Levin-CollectiveModeContributions-2017,Boyack-Lopes-ElectromagneticResponseSuperconductors-2020}.

    In this paper, we extend the theory developed in Ref.~\cite{Boyack-Boyack-RestoringGaugeInvariance-2018} to an ultraclean multicomponent superconductor.
    Specifically, we use the functional-integral formulation of the problem to derive the Gaussian-level fluctuation corrections to the electromagnetic linear response kernel.
    We pay special attention to the aspects of this analysis that are non-trivially altered by the multicomponent character of the superconducting order approached from the normal state.
    For more background on the theory of superconducting fluctuations we refer the reader to Refs.~\cite{Larkin-Varlamov-TheoryFluctuationsSuperconductors-2005,Varlamov-Glatz-FluctuationSpectroscopyRayleighJeans-2018}.
    
	The rest of this paper is structured as follows. 
	In Sec.~\ref{sec:effective_theory} we introduce the model for the multicomponent superconductor and construct its effective Gaussian fluctuation action. 
	In Sec.~\ref{sec:electro}, its electrodynamic response is computed, followed by an explicit demonstration of gauge invariance. 
	Sections \ref{sec:effective_theory} and \ref{sec:electro} are generalizations of the derivation by Boyack in Ref.~\cite{Boyack-Boyack-RestoringGaugeInvariance-2018} to the case of a multicomponent BCS superconductor.
	In Sec.~\ref{sec:fluct_con} we compute the fluctuation contribution to the electrical conductivity and discuss how the multicomponent character of the incipient superconducting order modifies the response.
	A summary with conclusions is presented in Sec.~\ref{sec:conc}.
	We use units in which $\hbar = k_{\mathrm{B}} = 1$.
	
	\section{Effective multicomponent fluctuation action}\label{sec:effective_theory}
	
	\subsection{Multicomponent BCS superconductor action}\label{sec:model}
	
	We consider the generalization of the BCS theory to the case of a Fermi liquid with multiple fermion ``flavors" \cite{Suhl-Walker-BardeenCooperSchriefferTheorySuperconductivity-1959}.
	Its second-quantized Hamiltonian is given by
	\begin{align}\label{eq:hamiltonian}
			\hat{H} &= \sum_{\sfa \sigma} \int \D^{d} r c_{\sigma \sfa}^{\dagger}(\v{r}) \Big( \varepsilon_{\sfa}(-\iu\bm{\nabla}) - \mu \Big) c_{\sigma\sfa}^{\mathstrut}(\v{r}) \\
			- &\sum_{\sfa\sfb} \int \D^{d} r \int \D^{d}r' g_{\sfa\sfb}(\v{r}-\v{r}') c^{\dagger}_{\uparrow\sfa}(\v{r}) c^{\dagger}_{\downarrow\sfa}(\v{r}') c_{\downarrow\sfb}^{\mathstrut}(\v{r}') c_{\uparrow\sfb}^{\mathstrut}(\v{r}). \notag
	\end{align}
	The operators $c_{\sigma\sfa}^{\phantom{\dagger}}$ and $c^{\dagger}_{\sigma\sfa}$ are the anticommuting annihilation and creation operators for an electron with spin $\sigma \in \{\uparrow,\downarrow\}$ and flavor $\sfa \in \{1,\dots,N\}$, where $N$ is the number of bands crossing the Fermi surface.
	We assume that the different flavors have parabolic dispersions $\varepsilon_{\sfa}(\v{k}) = \v{k}^2/(2m_{\sfa})$ with effective masses $m_{\sfa}$.
	The electron chemical potential is denoted by $\mu$.
	Following Ref.~\cite{Boyack-Boyack-RestoringGaugeInvariance-2018}, we assume the interaction potential to be a contact interaction $g_{\sfa\sfb}(\v{r}-\v{r}') = g_{\sfa\sfb} \delta(\v{r}-\v{r}')$, but its matrix structure now embodies the possibility of interband coupling \cite{Koshelev-Vinokur-TheoryFluctuationsTwoband-2005,Fanfarillo-Grilli-TheoryFluctuationConductivity-2009}.
	The interaction matrix $\mathbb{G} \coloneqq (g_{\sfa\sfb})$ is assumed to be real-valued and to have positive eigenvalues, denoted by $\lambda_{a}$. This puts restrictions on the strength of the interband coupling elements compared to the intraband coupling matrix elements, and in particular rules out the case of interband coupling only.  
	  $\mathbb{G}$ is diagonalized by the orthogonal matrix $S$ with entries $s_{a\sfa}$, i.e.,
	$ g_{\sfa\sfb} = \sum_{a}\lambda_{a} s_{a\sfa} s_{a\sfb}$.
	An illustration of the possible superconducting pairing processes allowed by the interaction potential $g_{\sfa\sfb}$ is shown in Fig.~\ref{fig:multiband_scattering}.
    Note that we do not account for pairing of electrons in different bands and we neglect other interaction channels such as those favoring a spin-density wave instability.

	The coherent-state functional-integral representation of the partition function is now introduced as
	\begin{subequations}\label{eq:S_fermions}
		\begin{align}
			Z &\coloneqq \int \DD \bar{\psi} \DD \psi \e^{-S[\bar{\psi},\psi]} \\
			\begin{split}
				S[\bar{\psi},\psi] &\coloneqq \int_{0}^{\beta} \D \tau \bigg(  \sum_{\sigma\sfa} \int \D^{d} r\, \bar{\psi}_{\sigma\sfa} (\tau,\v{r}) \partial_{\tau} \psi_{\sigma\sfa}(\tau,\v{r}) \\
				&\hspace{6em}+ H(\bar{\psi},\psi) \bigg),
			\end{split}
		\end{align}
	\end{subequations}
	where we have replaced the anticommuting operators on Fock space $c_{\sigma\sfa}^{\phantom{\dagger}}$ and $c^{\dagger}_{\sigma\sfa}$ by Grassmann-valued fields $\psi_{\sigma\sfa}$ and $\bar{\psi}_{\sigma\sfa}$ respectively.
	In the following, we will use the four-vector notation $x^{\mu} \coloneqq (t,\v{r}) \equiv (- \iu \tau, \v{r})$ and the shorthand notation $\int \D x \coloneqq \int \D \tau \int \D^{d} r$.
	We use the metric $\mathfrak{g} \coloneqq \mathrm{diag}(1,-1,\dots,-1)$.
	
	The bi-quadratic interaction between the fermions is eliminated in favor of a linear coupling between the fermion bilinears and an auxiliary boson through the Hubbard-Stratonovich decoupling in the particle-particle channel.
	This is done by introducing the $N$ auxiliary complex fields $\{\phi_{a}\}$ together with a functional integral measure normalized so that
	\begin{equation}
		1 = \int \DD[\bar{\phi},\phi] \exp( - \sum_{a} \int \D x \frac{ \bar{\phi}_{a}(x) \phi_{a}(x)}{\lambda_{a}}).
	\end{equation}
	Inserting this resolution of the identity into the partition function and performing the shift of the integration variables 
	\begin{equation}\label{eq:HS_shift}
		\phi_{a}(x) \mapsto \phi_{a}(x) - \sum_{\sfa} \lambda_a s_{a\sfa} \psi_{\downarrow \sfa}(x) \psi_{\uparrow \sfa}(x),
	\end{equation}
	yields
	\begin{align}\label{eq:Sint_diagonalized}
		S_{\mathrm{HS}}[\bar{\psi},\psi,\bar{\phi},\phi] &= \int \D x \sum_{\sigma \sfa} \bar{\psi}_{\sigma  \sfa }(x) \left( \partial_{\tau} + \xi_{\sfa}(-\iu \bm{\nabla}) \right) \psi_{\sigma \sfa}(x) \notag \\
		+ &\int \D x \sum_{a} \frac{\abs{\phi_{a}(x)}^2}{\lambda_{a}}  \\
		- &\int \D x \sum_{a \sfa } s_{a\sfa} \Big[ \bar{\phi}_{a} (x)  \psi_{\downarrow \sfa }(x) \psi_{\uparrow \sfa}(x)+ \cc \Big]. \notag
	\end{align}
    Note that the requirement that $\mathbb{G}$ be positive definite dictates the form of the shift in Eq.~\eqref{eq:HS_shift}. 
    Departing from this restriction necessitates a different Hubbard-Stratonovich decoupling \cite{Fanfarillo-Grilli-TheoryFluctuationConductivity-2009}.
    
	To facilitate integrating out the fermions, it is useful to write the action in the following form
	\begin{align}
			S_{\mathrm{HS}}[\bar{\psi},\psi,\bar{\phi},\phi] &= \sum_{\sfa}\int \D x \int \D y  \Psi_{\sfa}^{\dagger}(x) \left( - \mathcal{G}^{-1}_{\sfa}(x,y) \right) \Psi_{\sfa}(y) \notag \\
			&+ \sum_{a} \int \D x \frac{\abs{\phi_{a} (x)}^2}{\lambda_a},
	\end{align}
	where 
	\begin{subequations}
		\begin{equation}\label{eq:Greens-function}
			\begin{split}
				\mathcal{G}^{-1}_{\sfa}(x,y) &= \mathcal{G}^{-1}_{0,\sfa}(x,y) \\
				&+\sum_{a} s_{a \sfa} \left[ \phi_{a}(x) \sigma_{+} + \bar{\phi}_{a}(x) \sigma_{-} \right] \delta(x-y),
			\end{split}
		\end{equation}
		and
		\begin{equation}\label{eq:Free_Greens-function}
			\begin{split}
				\mathcal{G}^{-1}_{0,\sfa}(x,y) &= - \left[ \partial_{\tau} \sigma_{0} + \xi_{\sfa}(-\iu\bm{\nabla}) \sigma_{3}  \right] \delta(x-y) \\
				&= \begin{pmatrix}
					G_{0,\sfa}^{-1}(x,y) & \\
					& \tilde{G}_{0,\sfa}^{-1}(x,y)
				\end{pmatrix},
			\end{split}
		\end{equation}
	\end{subequations}
	where $G_{0,\sfa}(x,y)$ and $\tilde{G}_{0,\sfa}(x,y)$ denote the bare particle and hole propagators respectively \cite{Altland-Simons-CondensedMatterField-2023}.
	Here, the Pauli matrices $\sigma_{\mu}$ act on the Nambu spinor $\Psi_{\sfa}(x) \coloneqq \left( \psi_{\uparrow \sfa}(x) \quad \bar{\psi}_{\downarrow\sfa}(x) \right)^{\mathsf{T}}$, and $\sigma_{\pm} \coloneqq (\sigma_{1} \pm \iu \sigma_{2})/2$.
	Integrating out the fermions yields the effective bosonic theory
	\begin{equation}\label{eq:S_HS}
		\begin{split}
			S_{\mathrm{HS}}[\bar{\phi},\phi] = - \tr \log &\left( - \beta \mathcal{G}^{-1}[\bar{\phi},\phi] \right) \\
			&+ \sum_{a} \int \D x \frac{\abs{\phi_{a}(x)}^2}{\lambda_a}.
		\end{split}
	\end{equation}
	
	\subsection{Gaussian fluctuation action}

	Equation ~\eqref{eq:S_HS} is an exact rewriting of the original interacting fermionic theory in Eq.~\eqref{eq:S_fermions}.
	This rewriting is only computationally useful provided that an approximate treatment of the fluctuations in the auxiliary field $\phi$ describes the desired physics \cite{Kleinert-Kleinert-HubbardStratonovichTransformationSuccesses-2011}.
	Being concerned with fluctuation effects in the normal state but close to the superconducting transition, we can expand the action in Eq.~\eqref{eq:S_HS} in fluctuations about the trivial saddle point $\phi(x) = \phi_{\mathrm{mf}} + \eta(x) \equiv \eta(x)$ \cite{Boyack-Boyack-RestoringGaugeInvariance-2018}. 
	The resulting action yields the Ginzburg-Landau (GL) functional.
	In the immediate vicinity of the critical temperature, it is permitted to truncate the GL theory to quadratic order in the critical fluctuations $\eta$.
	Expanding the exponentiated fermion determinant to quadratic order yields (See Appendix~\ref{app:fluctuation_prop} for details)
	\begin{equation}\label{eq:S_eff}
		\begin{split}
			&S_{\mathrm{eff}}[\bar{\eta},\eta] = - \tr\log\left( -\beta \mathcal{G}^{-1}_{0} \right) \\
			&+\int \D x \int \D y \sum_{ab} \bar{\eta}_{a}(x) \left[ - L^{-1}(x-y) \right]_{ab} \eta_{b}(y),
		\end{split}
	\end{equation}
	where the inverse pair fluctuation propagator $L^{-1}(x-y)$ is given by
	\begin{equation}
		\begin{split}
			\left[- L^{-1}(x-y) \right]_{ab} &= \frac{\delta_{ab} \delta(x-y)}{\lambda_a} \\
			&+ \sum_{\sfa} s_{a\sfa} s_{b\sfa} G_{0,\sfa}(x,y) \tilde{G}_{0,\sfa}(y,x).
		\end{split}
	\end{equation} 
	In the Fourier-space representation the pair-fluctuation propagator reads
	\begin{equation}\label{eq:Linv}
		L^{-1}_{ab}(q) = - \frac{\delta_{ab}}{\lambda_a} + \left( S \Pi(q) S^{\mathsf{T}} \right)_{ab},
	\end{equation}
	where 
	\begin{equation}\label{eq:Pi}
		\Pi_{\sfa\sfb}(q) \coloneqq \delta_{\sfa\sfb} \frac{1}{\beta V} \sum_{k} G_{0,\sfa}(k) G_{0,\sfa}(q-k),
	\end{equation}
	denotes the particle-particle bubble.
	Here we are using the notation $q^{\mu} \coloneqq (q_{0},\v{q}) = (\iu\Omega_m,\v{q})$, and $\sum_{k} \coloneqq \sum_{\v{k}} \sum_{n\in\ZZ}$.
	
\begin{figure}[tb]
		\centering
		\includegraphics[width=\columnwidth]{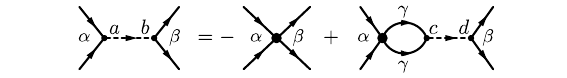}
		\caption{Diagrammatic representation of the Dyson equation for the fluctuation propagator $L_{ab}(q)$ \cite{Larkin-Varlamov-TheoryFluctuationsSuperconductors-2005,Koshelev-Vinokur-TheoryFluctuationsTwoband-2005}.
			The fermion propagator is illustrated with plain lines and the fluctuation propagator with dashed lines.
			The large circle represents the interaction matrix $g_{\sfa\sfb}$, while the small circle represents the coupling matrix $s_{a\sfa}$.
		}
		\label{fig:fluctuation_propagator}
	\end{figure}

	At this point, it is clear that the truncation of the GL functional at the Gaussian fluctuation level is equivalent to the ladder approximation of the Dyson equation for the pair fluctuation propagator \cite{Larkin-Varlamov-TheoryFluctuationsSuperconductors-2005}
	\begin{equation}
		(S^{\mathsf{T}} L(q) S )_{\sfa\sfb} = - g_{\sfa\sfb} + \sum_{\sfc\sfd} g_{\sfa\sfc} \Pi_{\sfc\sfd}(q)  (S^{\mathsf{T}} L(q) S )_{\sfd\sfb}.
	\end{equation}
	This equation is illustrated graphically in Fig.~\ref{fig:fluctuation_propagator}.

	\section{Electrodynamic response}\label{sec:electro}
	
	Electromagnetic linear response theory is the formalism used to find the linear functional relating the four-current $J^{\mu} = (\iu \rho, \v{J})$ to the electromagnetic gauge potential $A^{\mu} = (-\iu A_{0}, \v{A})$ \cite{Altland-Simons-CondensedMatterField-2023,Dupuis-Dupuis-FieldTheoryCondensed-2023}
	\begin{equation}\label{eq:lin_response_current}
		J^{\mu}(x) = \int \D x' \, K^{\mu\nu}(x,x') A_{\nu}(x'),
	\end{equation}
	which derives from the following effective Euclidean action
	\begin{equation}\label{eq:SeffA}
		S_{\mathrm{eff}}[A] = \frac{1}{2} \int \D x \int \D x' A_{\mu}(x) K^{\mu\nu}(x,x') A_{\nu}(x').
	\end{equation}
	This effective action is found by performing the minimal substitution $\partial_{\tau} \mapsto \partial_{\tau} - \iu e A_{0}$ and $-\iu\bm{\nabla} \mapsto - \iu\bm{\nabla} + e \v{A}$ in the bare fermion Green's functions appearing in the Gaussian fluctuation action in Eq.~\eqref{eq:S_eff} and subsequently integrating out the critical fluctuations $\eta$. 
    The minimal substitution does not alter the trivial saddle point, meaning that the identification of the critical fluctuations $\eta$ still holds in the presence of the external gauge potential.
    We note that this is not the case for the non-trivial saddle point below $T_{c}$ \cite{Anderson-Levin-GoingBCSLevel-2016,Boyack-Levin-CollectiveModeContributions-2017}.
	
	The linear response kernel appearing in Eq.~\eqref{eq:lin_response_current} is constrained by the principles of gauge invariance and charge conservation \cite{Altland-Simons-CondensedMatterField-2023,Schrieffer-Schrieffer-TheorySuperconductivity-1999}.
	Indeed, by performing a gauge transformation $A_{\mu} \mapsto A_{\mu} - \partial_{\mu} \chi$, the physical current is left invariant provided that $
		K^{\mu\nu}(x,x') \, \backvec{\partial}_{\nu}' = 0
	$.
	Moreover, the continuity equation $\partial_{\mu} J^{\mu}(x) = 0$ is satisfied if $\partial_{\mu} K^{\mu\nu}(x,x') = 0$.
	In a translationally invariant system, $K^{\mu\nu}(x,x') = K^{\mu\nu}(x-x')$ and these constraints are equivalent \cite{Altland-Simons-CondensedMatterField-2023}.
	In a free theory these constraints will be manifestly satisfied. 
	The issue is more subtle in interacting theories where the calculation of  $K^{\mu\nu}(x,x')$ is necessarily approximate \cite{Schrieffer-Schrieffer-TheorySuperconductivity-1999}. 
	In general, the problem can be solved by making only approximations consistent with the generalized Ward identity \cite{Schrieffer-Schrieffer-TheorySuperconductivity-1999,Watanabe-Watanabe-GaugeinvariantElectromagneticResponses-2025}.
    These approximations are known as $\Phi$-derivable, and come from the two-particle irreducible effective action formalism \cite{Luttinger-Ward-GroundStateEnergyManyFermion-1960,Baym-Kadanoff-ConservationLawsCorrelation-1961,Baym-Baym-SelfConsistentApproximationsManyBody-1962}. However, as shown in Ref.~\cite{Boyack-Boyack-RestoringGaugeInvariance-2018}, the Gaussian fluctuation approximation is exactly one such approximation, giving rise to a consistent set of diagrams for the fluctuation corrections of $K^{\mu\nu}$.
	When computing transport properties in the critical regime, we therefore follow this approach.
	
	\subsection{Fluctuation response kernel}\label{sec:response_kernel}

    \begin{figure*}[t]
		\centering
		\captionsetup{skip=0pt}
		\includegraphics[width=\textwidth]{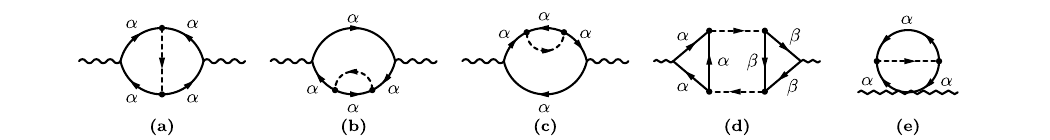}
		\caption{Fluctuation corrections to the electromagnetic response kernel with fermion band labels included. Panel~\textbf{(a)} is the Maki-Thompson (MT) diagram, panels~\textbf{(b)} and \textbf{(c)} are the density-of-states (DOS) diagrams, panel~\textbf{(d)} is the Aslamazov-Larkin (AL) diagram and panel~\textbf{(e)} is the diamagnetic (DIA) diagram.}
		\label{fig:fluctuation_diagrams}
	\end{figure*}	
	
	Following Ref.~\cite{Boyack-Boyack-RestoringGaugeInvariance-2018}, we focus on the fluctuation correction to the response kernel
	\begin{equation}
		K^{\mu\nu}_{\mathrm{fluc}}(x,x') = \frac{\delta^2 \tr\log\left(-\beta L^{-1}[A]\right)}{\delta A_{\mu}(x) \delta A_{\nu}(x')} \biggr\lvert_{A = 0}. 
	\end{equation}
	Performing the two functional derivatives yields (See Appendix~\ref{app:fluctuation_response} for details)
	\begin{widetext}
		\begin{subequations}
			\begin{align}
				K^{\mu\nu}_{\mathrm{fluc}}(x,x') = &- \int \prod_{i=1}^{4} \D y_{i} \tr \left[L(y_1,y_2) \Lambda^{\mu}(y_2,y_3;x) L(y_3,y_4) \Lambda^{\nu}(y_4,y_1;x') \right] \label{eq:K_fluctuation_threept} \\
				&+ \int \prod_{i=1}^{2} \D y_{i} \tr\left[ L(y_1,y_2) \Gamma^{\mu\nu}(y_2,y_1;x,x') \right], \label{eq:K_fluctuation_fourpt}
			\end{align}
		\end{subequations}
	\end{widetext}
	where we employ a similar notation for the effective vertices to that in Ref.~\cite{Boyack-Boyack-RestoringGaugeInvariance-2018}
	\begin{subequations}
		\begin{align}
			\Lambda^{\mu}(y_1,y_2;x) &\coloneqq \frac{\delta L^{-1}[A](y_1,y_2)}{\delta A_{\mu}(x)} \biggr\lvert_{A = 0}, \label{eq:threept_vertex} \\
			\Gamma^{\mu\nu}(y_1,y_2;x,x') &\coloneqq \frac{\delta^2 L^{-1}[A](y_1,y_2)}{\delta A_{\mu}(x) \delta A_{\nu}(x')} \biggr\lvert_{A = 0}. \label{eq:fourpt_vertex}
		\end{align}
	\end{subequations}
	At this point, we note that the only principal difference between this result and the single-component counterpart in Ref.~\cite{Boyack-Boyack-RestoringGaugeInvariance-2018} is the presence of the remaining matrix trace. 
	As will become evident in the following, this is precisely what gives rise to the unique transport signatures of the incipient multicomponent superconducting order.
	
	The correction to the response kernel in Eq.~\eqref{eq:K_fluctuation_threept} can be interpreted as a particle-hole bubble of pair-fluctuation propagators, with a current vertex given by Eq.~\eqref{eq:threept_vertex} inherited from that of the electrons.
	This is the Aslamazov-Larkin (AL) diagram \cite{Aslamazov-Larkin-EffectFluctuationsProperties-1968}.
	The correction in Eq.~\eqref{eq:K_fluctuation_fourpt} contains the density-of-states (DOS) diagrams, the Maki-Thompson (MT) diagram and the diamagnetic (DIA) diagram \cite{Boyack-Boyack-RestoringGaugeInvariance-2018}.
	All of these can be assigned interpretations as interaction-dressed versions of the bare diagrams contributing to the electromagnetic response kernel.
	The diagrams are illustrated graphically in Fig.~\ref{fig:fluctuation_diagrams}.
	That these diagrams should be treated on equal footing is not apparent from the diagrammatic quantum field theory point of view. 
	The added appeal from the functional-integral derivation advocated by Ref.~\cite{Boyack-Boyack-RestoringGaugeInvariance-2018} is the demonstration that this set of diagrams does not derive from a perturbative expansion. 
	Rather, it is the complete set of diagrams to be considered in linear response theory under the assumption of Gaussian fluctuations in the particle-particle channel.

	In Appendix~\ref{app:fluctuation_response} we derive the diagrams shown in Fig.~\ref{fig:fluctuation_diagrams} from the effective action by performing the functional derivatives with respect to the external gauge field.
	Here, we write down the resulting expressions in momentum space.
	The Fourier-space representation of Eq.~\eqref{eq:SeffA} in a translationally invariant system is 
	\begin{equation}
		S_{\mathrm{eff}}[A] = \frac{1}{2}\sum_{q} K^{\mu\nu}(q) A_{\mu}(-q) A_{\nu}(q),
	\end{equation}
	where 
	$
		K^{\mu\nu}(q) \coloneqq \int \D x \, \e^{-\iu q \cdot x} K^{\mu\nu}(x).
	$
	The fluctuation correction to the response kernel is decomposed as
	\begin{equation}
		K^{\mu\nu}_{\mathrm{fluc}}(q) = K^{\mu\nu}_{\mathrm{MT}}(q) + K^{\mu\nu}_{\mathrm{DOS}}(q)+ K^{\mu\nu}_{\mathrm{AL}}(q) + K^{\mu\nu}_{\mathrm{DIA}}(q),
	\end{equation}
	where
	\begin{widetext}
		\begin{subequations}\label{eq:diagram_expressions}
			\begin{align}
				\begin{split}
					K^{\mu\nu}_{\mathrm{MT}}(q) &= \frac{2e^2 }{(\beta V)^2} \sum_{k p}  \sum_{\sfa} \left( S^{\mathsf{T}} L(p) S \right)_{\sfa\sfa}  G_{0,\sfa}(k-q) \gamma_{\sfa}^{\mu}(k-q,k)  G_{0,\sfa}(k) \\
					&\hspace{12.1em} \times G_{0,\sfa}(p-k) \gamma_{\sfa}^{\nu}(p+q-k,p-k) G_{0,\sfa}(p+q-k) \phantom{\sum_{k}}
				\end{split}\\
				\begin{split}
					K^{\mu\nu}_{\mathrm{DOS}}(q)&=  \frac{2 e^2}{(\beta V)^2} \sum_{k p}  \sum_{\sfa} \left( S^{\mathsf{T}} L(p) S \right)_{\sfa\sfa} G_{0,\sfa}(k-q) \gamma^{\mu}_{\sfa}(k-q,k) G_{0,\sfa}(k)  G_{0,\sfa}(p-k) G_{0,\sfa}(k) \gamma^{\nu}_{\sfa}(k,k-q) \\
					&\,+ \frac{ 2 e^2}{(\beta V)^2} \sum_{k p} \sum_{\sfa} \left( S^{\mathsf{T}} L(p) S \right)_{\sfa\sfa} G_{0,\sfa} (k) \gamma^{\mu}_{\sfa}(k,k+q) G_{0,\sfa}(k+q) \gamma_{\sfa}^{\nu}(k+q,k) G_{0,\sfa}(k) G_{0,\sfa}(p-k) 
				\end{split} \\
				\begin{split}
					K^{\mu\nu}_{\mathrm{AL}}(q) &= - \frac{4e^2}{(\beta V)^3} \sum_{k k' p} \sum_{\sfa\sfb} \left(S^{\mathsf{T}} L(p-q) S \right)_{\sfa\sfb} \left( S^{\mathsf{T}} L(p) S \right)_{\sfb\sfa} G_{0,\sfa}(k-q) \gamma^{\mu}_{\sfa}(k-q,k) G_{0,\sfa}(k)  G_{0,\sfa}(p-k) \\
					&\hspace{20.5em}\times G_{0,\sfb}(k') \gamma^{\nu}_{\sfb}(k',k'-q) G_{0,\sfb}(k'-q) G_{0,\sfb}(p-k') \phantom{\sum_{k}} 
				\end{split}\\
				K^{\mu\nu}_{\mathrm{DIA}}(q) &= \frac{2e^2}{(\beta V)^2} \sum_{k p }  \sum_{\sfa} \frac{1}{m_{\sfa}} \delta^{\mu\nu}(1 - \delta^{\mu 0}) \left( S^{\mathsf{T}} L(p) S \right)_{\sfa\sfa} \left[ G_{0,\sfa}(k) \right]^2 G_{0,\sfa}(p-k).
			\end{align}
		\end{subequations}
		\end{widetext}
        In these expressions, we have introduced the bare particle current vertex $\gamma^{\mu}_{\sfa}(k+q,k)$, which is related to the paramagnetic current as $j_{\mathrm{p}}^{\mu}(q) = - e \sum_{k} \sum_{\sigma \sfa} \gamma^{\mu}_{\sfa}(k+q,k) \bar{\psi}_{\sigma \sfa}(k) \psi_{\sigma \sfa}(k+q)$ (See Appendix~\ref{app:fluctuation_response} for details). 
        
        From the expressions in Eq.~\eqref{eq:diagram_expressions}, as well as their diagrammatic representations in Fig.~\ref{fig:fluctuation_diagrams}, one observes that the AL diagram stands out as special because it contains two pair-fluctuation propagators rather than one. 
        These propagators become critical at the transition temperature, and this diagram is therefore expected to make the largest contribution to the response kernel in the critical regime \cite{Aslamazov-Larkin-EffectFluctuationsProperties-1968}.
        However, note that it also contains two more electron Green's functions.
        Given that the inverse pair-fluctuation propagator contains the particle-particle bubble, it is plausible that the presence of the extra pair-fluctuation propagator is compensated by these two electron Green's functions, ultimately making the AL diagram comparable to the DOS and MT diagrams.
        In Sec.~\ref{sec:fluct_con}, we prove that this is indeed the case.
        
		\subsection{Ward identity}\label{sec:gauge_invariance}
		
		To demonstrate the consistency of the Gaussian-fluctuation approximation, we now aim to show that the fluctuation kernel satisfies $q_{\mu} K^{\mu\nu}(q) = 0$.
		This was done for the single-component case in Ref.~\cite{Boyack-Boyack-RestoringGaugeInvariance-2018} and here we present its multicomponent generalization.
		
		The basic tool needed to demonstrate gauge invariance is the \textit{bare} Ward-Takahashi identity given by 
		\begin{equation}
			q_{\mu} \gamma^{\mu}_{\sfa}(k+q,k) =  G_{0,\sfa}^{-1}(k+q) - G_{0,\sfa}^{-1}(k),
		\end{equation}
		where the bare current vertex has components $\gamma^{0}_{\sfa}(k+q,k) = 1$ and $\gamma^{i}_{\sfa}(k+q,k) = (k+q/2)^i/m_{\sfa}$.
		Let us first do the contraction with $K^{\mu\nu}_{\mathrm{AL}}(q)$, which is the most complicated of the diagrams.
		This yields 
		\begin{widetext}
		\begin{align}
			\begin{split}
				q_{\mu} K^{\mu\nu}_{\mathrm{AL}}(q) = - \frac{4e^2}{(\beta V)^3} \sum_{k k' p } \sum_{\sfa\sfb} \left(S^{\mathsf{T}} L(p-q) S \right)_{\sfa\sfb} \left( S^{\mathsf{T}} L(p) S \right)_{\sfb\sfa} &  G_{0,\sfa}(k-q) \left[ G_{0,\sfa}^{-1}(k) - G_{0,\sfa}^{-1}(k-q) \right] G_{0,\sfa}(k) G_{0,\sfa}(p-k) \\
				\times & \, G_{0,\sfb}(k') \gamma^{\nu}_{\sfb}(k',k'-q) G_{0,\sfb}(k'-q) G_{0,\sfb}(p-k') 
			\end{split} \notag \\
			\begin{split}
				=- \frac{4e^2}{(\beta V)^2} \sum_{k' p } \sum_{\sfa\sfb} \left(S^{\mathsf{T}} L(p-q) S \right)_{\sfa\sfb} \left( S^{\mathsf{T}} L(p) S \right)_{\sfb\sfa}\frac{1}{\beta V} \sum_{k}&\Big[ G_{0,\sfa}(k-q) G_{0,\sfa}(p-k) - G_{0,\sfa}(k) G_{0,\sfa}(p-k) \Big] \\
				&\hspace{-0.75em} \times G_{0,\sfb}(k') \gamma^{\nu}_{\sfb}(k',k'-q) G_{0,\sfb}(k'-q) G_{0,\sfb}(p-k').
				\end{split}
		\end{align}
		Next, we note that the integrated quantity inside the square brackets above corresponds to a difference of two particle-particle bubbles, as defined by Eq.~\eqref{eq:Pi}.
		Using Eq.~\eqref{eq:Linv} we can write this as
		\begin{align}
			\begin{split}
				q_{\mu} K^{\mu\nu}_{\mathrm{AL}}(q) = - \frac{4e^2}{(\beta V)^2} \sum_{k' p} \sum_{\sfa\sfb\sfc} &\left(S^{\mathsf{T}} L(p-q) S \right)_{\sfc\sfb} \left( S^{\mathsf{T}} L(p) S \right)_{\sfb\sfa} \left[ S^{\mathsf{T}} \left( L^{-1}(p-q) - L^{-1}(p)  \right) S \right]_{\sfa\sfc} \\
				&\hspace{10.5em}\times G_{0,\sfb}(k') \gamma^{\nu}_{\sfb}(k',k'-q) G_{0,\sfb}(k'-q) G_{0,\sfb}(p-k')
			\end{split} \\
			= - \frac{4e^2}{(\beta V)^2} \sum_{k' p} \sum_{\sfa}& \left[ S^{\mathsf{T}} \left( L(p) - L(p-q) \right) S \right]_{\sfa\sfa} G_{0,\sfa}(k') \gamma^{\nu}_{\sfa}(k',k'-q) G_{0,\sfa}(k'-q) G_{0,\sfa}(p-k') \\
			= + \frac{4e^2}{(\beta V)^2} \sum_{k' p} \sum_{\sfa}& (S^{\mathsf{T}}L(p)S)_{\sfa\sfa} \Big[ G_{0,\sfa}(p+q-k') - G_{0,\sfa}(p-k') \Big] G_{0,\sfa}(k') \gamma^{\nu}_{\sfa}(k',k'-q) G_{0,\sfa}(k'-q).
		\end{align}
		Here, we have made repeated use of the fact that $\sum_{c} L^{-1}_{ac}(p) L_{cb}(p) = \delta_{ab}$, the orthogonality of the matrix $S$, and performed shifts of the momenta so that the contraction of the AL diagram appears in a form similar to that of the remaining diagrams:
		\begin{align}
			q_{\mu} K^{\mu\nu}_{\mathrm{MT}}(q) &= \frac{2e^2}{(\beta V)^2} \sum_{kp} \sum_{\sfa} (S^{\mathsf{T}}L(p) S)_{\sfa\sfa} \Big[ G_{0,\sfa}(k-q) - G_{0,\sfa}(k) \Big] G_{0,\sfa}(p-k) G_{0,\sfa}(p-k+q) \gamma_{\sfa}^{\nu}(p-k+q,p-k) \\
			\begin{split}
				q_{\mu} K^{\mu\nu}_{\mathrm{DOS}}(q) &= \frac{2e^2}{(\beta V)^2} \sum_{kp} \sum_{\sfa}  (S^{\mathsf{T}}L(p) S)_{\sfa\sfa} \Big[ G_{0,\sfa}(k-q) - G_{0,\sfa}(k) \Big] \gamma_{\sfa}^{\nu}(k,k-q) \\
				&\hspace{11.6em}\times \Big[ G_{0,\sfa}(p-k)G_{0,\sfa}(k) + G_{0,\sfa}(p-k+q) G_{0,\sfa}(k-q) \Big]
			\end{split} \\
			q_{\mu} K^{\mu\nu}_{\mathrm{DIA}}(q) &= \frac{2e^2}{(\beta V)^2} \sum_{kp} \sum_{\sfa} \frac{q_{\nu}}{m_{\sfa}}(1 - \delta^{\nu0}) (S^{\mathsf{T}}L(p) S)_{\sfa\sfa} \left[G_{0,\sfa}(k)\right]^2 G_{0,\sfa}(p-k).
		\end{align}
		Having demonstrated that the contracted fluctuation kernel takes on a ``diagonal form" in the band indices, the remainder of the proof follows exactly the same procedure as in Ref.~\cite{Boyack-Boyack-RestoringGaugeInvariance-2018}.
		Indeed, after some algebra, one finds that 
		\begin{align}
			q_{\mu} K^{\mu\nu}_{\mathrm{AL}}(q) + q_{\mu} K^{\mu\nu}_{\mathrm{MT}}(q) + q_{\mu} K^{\mu\nu}_{\mathrm{DOS}}(q) &= \frac{2e^2}{(\beta V)^2} \sum_{kp \sfa} (S^{\mathsf{T}}L(p) S)_{\sfa\sfa} \left[G_{0,\sfa}(k)\right]^2 G_{0,\sfa}(p-k) \left[ \gamma_{\sfa}^{\nu}(k+q,k) - \gamma^{\nu}_{\sfa}(k,k-q) \right] \notag \\
			&= \frac{2e^2}{(\beta V)^2} \sum_{kp \sfa} (S^{\mathsf{T}}L(p) S)_{\sfa\sfa} \left[G_{0,\sfa}(k)\right]^2 G_{0,\sfa}(p-k) \frac{q^{\nu}}{m_{\sfa}}\left( 1 - \delta^{\nu 0}\right),
		\end{align}
		which explicitly shows that $q_{\mu}K^{\mu\nu}_{\mathrm{fluc}}(q)=0$, as advertised.
\end{widetext}
	
\section{Fluctuation conductivity}\label{sec:fluct_con}
	
	Having demonstrated the soundness of the approximate calculation of $K^{\mu\nu}_{\mathrm{fluc}}$, we now turn to calculating its contribution to the optical conductivity.    
	This calculation has not, to our knowledge, been carried out for an ultraclean multicomponent superconductor.
	
	Working in the Weyl gauge $\iu A_{0} = 0$, the electric field is $\v{E}(\v{r},t) = -\partial_{t} \v{A}(\v{r},t)$ and the real part of the uniform conductivity tensor is given by 
	\begin{equation}
		\Re\left[\sigma^{ij}(\omega)\right] = \Re\left[ \frac{1}{\Omega_m} K^{ij}(\iu\Omega_m,\v{0})\biggr\lvert_{\iu\Omega_m\to \omega + \iu 0} \right].
	\end{equation}
	Using the Dirac identity $ 1/(x+\iu0) = \mathrm{P}(1/x) - \iu \pi \delta(x)$ one can express the real part of the longitudinal conductivity tensor as
	\begin{equation}
		\Re \sigma^{\parallel}(\omega) = D\delta(\omega) + \Re \sigma_{\mathrm{reg}}^{\parallel}(\omega),
	\end{equation}
	where $D$ denotes the Drude weight, and $\sigma_{\mathrm{reg}}(\omega)$ is the regular part of the conductivity tensor.
	The Drude weight receives a correction due to the fluctuations given by
	\begin{equation}
		\delta D = \lim_{\omega \to 0} \pi \Re [ K^{ii}_{\mathrm{fluc}}(\omega + \iu 0, \v{0}) ],
	\end{equation}
	while the regular (or dissipative) part of the conductivity in an ultraclean metal is determined solely by its fluctuation contribution and is given by
	\begin{equation}\label{eq:sigma_reg_def}
		\Re \sigma_{\mathrm{reg}}^{\parallel}(\omega) = - \mathrm{P}\frac{1}{\omega} \Im\left[ K^{ii}_{\mathrm{fluc}}(\omega + \iu 0,\v{0}) \right].
	\end{equation}
	In Ref.~\cite{Boyack-Boyack-RestoringGaugeInvariance-2018} the focus was on the Drude weight.
	Here we focus mainly on the regular conductivity.
	
	When computing the contraction between the fluctuation kernel and the external momenta in Sec.~\ref{sec:gauge_invariance}, we found that the result essentially reduced to a sum over single-band contributions.
	For the conductivity tensor, we will see shortly that such a simplification does not take place.
	In particular, this means that the AL diagram is genuinely different from the other diagrams, in the sense that it involves a summation over two band indices.
	Our strategy for simplifying the expression for the regular longitudinal conductivity will be to express the MT and DOS diagrams in a form comparable to the AL diagram.
	The DIA diagram is purely real and therefore does not contribute to the regular part of the conductivity tensor.
    Its effects on the Drude weight are analogous to those found by Ref.~\cite{Boyack-Boyack-RestoringGaugeInvariance-2018} and they will be discussed in Sec.~\ref{sec:sum_rule}.
	
	\subsection{AL diagram and triangle block}\label{sec:triangle}
	
\begin{figure}[htb]
		\centering
		\vspace{1em}
		\includegraphics[width=0.55\columnwidth]{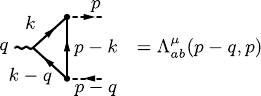}
		\caption{Triangle vertex block appearing in the AL diagram.}
		\label{fig:AL_vertex}
	\end{figure}

	Let us begin by simplifying the expression for the triangle block constituting the effective electromagnetic vertices for the pair fluctuations, shown in Fig.~\ref{fig:AL_vertex}.
	
	The Fourier representation of the vertex in Fig.~\ref{fig:AL_vertex} is found by Fourier transforming the expression for the vertex in Eq.~\eqref{eq:threept_vertex} derived in Appendix \ref{app:fluctuation_response} [See Eq.~\eqref{eq:Lambda_mu}], and it reads 
	\begin{equation}
		\begin{split}
			\Lambda^{\mu}_{ab}(p-q,p) &= -\frac{2e}{\beta V} \sum_{k} \sum_{\sfa} s_{a\sfa} s_{b\sfa} \gamma^{\mu}_{\sfa}(k-q,k)  \\
			 &\phantom{\sum_{k}} \times G_{0,\sfa}(k-q) G_{0,\sfa}(k) G_{0,\sfa}(p-k).
		\end{split}
	\end{equation}
	We evaluate this diagram in the uniform (long-wavelength) limit $\v{q} = 0$ and for $\mu=i$. 
    Taking the $\v{q}\to 0$ limit first in the computation of the conductivity tensor importantly implies that we are probing an infinite system. This ensures that energy levels can be arbitrary close, allowing for absorption even for $\omega \to 0$, which is a requirement of a metallic state \cite{Dupuis-Dupuis-FieldTheoryCondensed-2023}.
	For $\v{q}\to 0$ we can make use of the fact that $G_{0,\sfa}^{-1}(k) - G_{0,\sfa}^{-1}(k-q_{0}) = q_{0}$, where $k-q_{0}$ is to be understood as the four-vector $(k_0 - q_0, \v{k})$.
    Equivalently,
	\begin{equation}\label{eq:q_0_identity}
		G_{0,\sfa}(k) G_{0,\sfa}(k-q_{0}) = \frac{1}{q_{0}} \left[ G_{0,\sfa}(k-q_{0}) - G_{0,\sfa}(k) \right].
	\end{equation}
	Together with the fact that $\gamma^{\mu}_{\sfa}(k+q_{0},k) = \gamma^{\mu}_{\sfa}(k,k)$ this allows us to write 
	\begin{widetext}
		\begin{equation}
			\Lambda^{i}_{ab}(p-q_{0},p) = - \frac{1}{\beta V} \frac{2e}{q_{0}} \sum_{k \sfa} s_{a\sfa} s_{b\sfa} \gamma^{i}_{\sfa}(k,k) \bigg[G_{0,\sfa}(k-q_{0}) G_{0,\sfa}(p-k) -G_{0,\sfa}(k) G_{0,\sfa}(p-k)\bigg].
		\end{equation}
		Now, note that 
		\begin{align}
			\frac{1}{\beta V} \sum_{k} \gamma_{\sfa}^{i}(k,k) G_{0,\sfa}(p-k) G_{0,\sfa}(k-q_{0}) &= \frac{1}{\beta V}\sum_{k} \gamma_{\sfa}^{i}(-k+p,-k+p) G_{0,\sfa}(k) G_{0,\sfa}(-k-q_{0}+p) \label{eq:vertex_bubble_sum} \\
			\begin{split}
				&= - \frac{1}{\beta V} \sum_{k}  \gamma_{\sfa}^{i}(k,k) G_{0,\sfa}(k) G_{0,\sfa}(-k-q_{0}+p) \\
				&\hspace{1.1em}+ \frac{1}{\beta V } \sum_{k} \gamma_{\sfa}^{i}(p,p) G_{0,\sfa}(k) G_{0,\sfa}(-k-q_{0}+p) 
			\end{split} \\
				&=  \frac{1}{2} \gamma_{\sfa}^{i}(p,p) \frac{1}{\beta V} \sum_{k} G_{0,\sfa}(k) G_{0,\sfa}(-k-q_{0}+p).
		\end{align}
		where we have used that $\gamma^{i}_{\sfa}(-k+p,-k+p) = -\gamma^{i}_{\sfa}(k,k) + \gamma^{i}_{\sfa}(p,p)$.
		We recognize the remaining momentum sum as the particle-particle bubble of Eq.~\eqref{eq:Pi}. 
		Using Eq.~\eqref{eq:Linv} again we find in total
		\begin{align}\label{eq:vertex_uniform_limit}
			\Lambda^{i}_{ab} (p-q_{0},p) = \frac{e}{q_{0}} \sum_{\sfa} \gamma_{\sfa}^{i}(p,p) s_{a\sfa} s_{b\sfa} \left[ \left(S^{\mathsf{T}}  L^{-1}(p) S \right) - \left( S^{\mathsf{T}} L^{-1}(p-q_{0})  S \right) \right]_{\sfa\sfa} \equiv e \sum_{\sfa} \gamma^{i}_{\sfa}(p,p) s_{a\sfa} s_{b\sfa} C_{\sfa}(p,q_0), 
		\end{align}
		where we have introduced the shorthand notation $C_{\sfa}(p,q_{0})$ for the finite difference derivative of the inverse pair fluctuation propagator with respect to $p_0$.
		
		Using the expression for the vertex in Eq.~\eqref{eq:vertex_uniform_limit} we can write down the AL diagram in the uniform limit as
		\begin{equation}\label{eq:AL_uniform_limit}
			K^{ii}_{\mathrm{AL}}(q_0,\v{0}) = -e^2 \frac{1}{\beta V} \sum_{ p } 	\sum_{\sfa\sfb} \gamma_{\sfa}^{i}(p,p) \gamma_{\sfb}^{i}(p,p) C_{\sfa}(p,q_{0}) C_{\sfb}(p,q_{0}) \left( S^{\mathsf{T}} L(p) S \right)_{\sfb\sfa} \left( S^{\mathsf{T}} L(p-q_{0}) S \right)_{\sfa\sfb}.
		\end{equation}
		
		\subsection{MT and DOS diagrams}
		
		Let us now consider the MT diagram in the uniform limit $\v{q} = 0$.
		By making use of the identity in Eq.~\eqref{eq:q_0_identity} twice we can write 
		\begin{align}
			\begin{split}
				K^{ii}_{\mathrm{MT}} (q_0,\v{0}) &= \frac{2e^2}{(\beta V)^2} \sum_{kp} \sum_{\sfa} \left( S^{\mathsf{T}} L(p) S \right)_{\sfa\sfa} \gamma_{\sfa}^{i}(k,k) \gamma_{\sfa}^{i}(p-k,p-k) \\
				&\hspace{10em}\times \frac{1}{q_{0}} \left[ G_{0,\sfa}(k-q_{0}) - G_{0,\sfa}(k)  \right]  \frac{1}{q_{0}}\left[G_{0,\sfa}(p-k) - G_{0,\sfa}(p+q_{0}-k)\right]
            \end{split}\\
            \begin{split}
				&= \frac{2e^2}{q_{0}^2} \frac{1}{(\beta V)^2} \sum_{kp} \sum_{\sfa} \gamma_{\sfa}^{i}(k,k) \gamma_{\sfa}^{i}(p-k,p-k) G_{0,\sfa}(k) G_{0,\sfa}(p-k) \\
				&\hspace{10em}\times\left[ \left(S^{\mathsf{T}} L(p+q_{0}) S\right) + \left(S^{\mathsf{T}} L(p-q_{0}) S\right) - 2 \left(S^{\mathsf{T}} L(p) S\right) \right]_{\sfa\sfa}.
			\end{split}
		\end{align}
		Using again the fact that $\gamma^{i}_{\sfa}(p-k,p-k) = -\gamma^{i}_{\sfa}(k,k) + \gamma^{i}_{\sfa}(p,p)$ permits us to write  
		\begin{align}
			\begin{split}
				K^{ii}_{\mathrm{MT}} (q_0,\v{0}) &= -\frac{2e^2}{q_{0}^2} \frac{1}{(\beta V)^2} \sum_{kp} \sum_{\sfa} \left[ \gamma_{\sfa}^{i}(k,k)\right]^2 G_{0,\sfa}(k) G_{0,\sfa}(p-k) \\
				&\hspace{10em}\times\left[ \left(S^{\mathsf{T}} L(p+q_{0}) S\right) + \left(S^{\mathsf{T}} L(p-q_{0}) S\right) - 2 \left(S^{\mathsf{T}} L(p) S\right) \right]_{\sfa\sfa} \\
				&\phantom{=} \, + \frac{2e^2}{q_{0}^2} \frac{1}{(\beta V)^2} \sum_{kp} \sum_{\sfa}  \gamma_{\sfa}^{i}(k,k) \gamma_{\sfa}^{i}(p,p) G_{0,\sfa}(k) G_{0,\sfa}(p-k) \\
				&\hspace{10em}\times\left[ \left(S^{\mathsf{T}} L(p+q_{0}) S\right) + \left(S^{\mathsf{T}} L(p-q_{0}) S\right) - 2 \left(S^{\mathsf{T}} L(p) S\right) \right]_{\sfa\sfa}.
			\end{split}
		\end{align}
		At this point, we see that the summation over $k$ in the second term is of the same form as in Eq.~\eqref{eq:vertex_bubble_sum} with $q_{0} = 0$.
		This allows us to finally write
		\begin{subequations}
			\begin{align}
				\begin{split}
					K^{ii}_{\mathrm{MT}} (q_0,\v{0}) &= -\frac{2e^2}{q_{0}^2} \frac{1}{(\beta V)^2} \sum_{kp} \sum_{\sfa} \left[ \gamma_{\sfa}^{i}(k,k)\right]^2 G_{0,\sfa}(k) G_{0,\sfa}(p-k) \\
					&\hspace{10em}\times\left[ \left(S^{\mathsf{T}} L(p+q_{0}) S\right) + \left(S^{\mathsf{T}} L(p-q_{0}) S\right) - 2 \left(S^{\mathsf{T}} L(p) S\right) \right]_{\sfa\sfa} \label{eq:MT_uniform_1}
				\end{split}\\
				\begin{split}
					&\phantom{=} \, + \frac{e^2}{q_{0}^2} \frac{1}{\beta V} \sum_{p} \sum_{\sfa} \left[ \gamma_{\sfa}^{i}(p,p) \right]^2 \Pi_{\sfa}(p) \left[ \left(S^{\mathsf{T}} L(p+q_{0}) S\right) + \left(S^{\mathsf{T}} L(p-q_{0}) S\right) - 2 \left(S^{\mathsf{T}} L(p) S\right) \right]_{\sfa\sfa}. \label{eq:MT_uniform_2}
				\end{split}
			\end{align}
		\end{subequations}
		
		Let us now do a similar simplification of the DOS diagrams. 
		Repeated application of the identity in Eq.~\eqref{eq:q_0_identity} yields
		\begin{align}
			K^{ii}_{\mathrm{DOS}}(q_0,\v{0}) &= \frac{2e^2}{(\beta V)^2} \sum_{kp}\sum_{\sfa} (S^{\mathsf{T}} L(p) S)_{\sfa\sfa} \left[\gamma_{\sfa}^{i}(k,k)\right]^2 G_{0,\sfa}(k) G_{0,\sfa}(p-k) \frac{1}{q_{0}} \left[ G_{0,\sfa}(k-q_{0}) - G_{0,\sfa}(k+q_{0}) \right] \\
			&=  \frac{2e^2}{(\beta V)^2} \sum_{kp}\sum_{\sfa} (S^{\mathsf{T}} L(p) S)_{\sfa\sfa} \left[\gamma_{\sfa}^{i}(k,k)\right]^2 G_{0,\sfa}(p-k) \frac{1}{q_{0}^2} \left[ G_{0,\sfa}(k-q_{0}) + G_{0,\sfa}(k+q_{0}) - 2 G_{0,\sfa}(k) \right].
 		\end{align}
 		After some rearrangements, we find
 		\begin{equation}\label{eq:DOS_uniform}
 			\begin{split}
 				K^{ii}_{\mathrm{DOS}}(q_0,\v{0}) &= \frac{2e^2}{(\beta V)^2} \sum_{kp}\sum_{\sfa}  \left[\gamma_{\sfa}^{i}(k,k)\right]^2 G_{0,\sfa}(k) G_{0,\sfa}(p-k) \\
 				&\hspace{10em} \times \frac{1}{q_{0}^2} \left[ \left(S^{\mathsf{T}} L(p+q_{0}) S\right) + \left(S^{\mathsf{T}} L(p-q_{0}) S\right) - 2 \left(S^{\mathsf{T}} L(p) S\right) \right]_{\sfa\sfa}.
 			\end{split}
 		\end{equation}
 		At this point, we realize that the first term of the MT diagram in the uniform limit in Eq.~\eqref{eq:MT_uniform_1} cancels exactly against the DOS diagram in Eq.~\eqref{eq:DOS_uniform}.
		This leaves only the second term of the MT diagram in Eq.~\eqref{eq:MT_uniform_2}.
		This can in turn be rearranged slightly and brought to the following form
		\begin{align}
			K^{ii}_{\mathrm{MT}}(q_0,\v{0})  + K^{ii}_{\mathrm{DOS}}(q_0,\v{0}) &= \frac{e^2}{q_{0}^2} \frac{1}{\beta V} \sum_{p} \sum_{\sfa}  \left[ \gamma_{\sfa}^{i}(p,p) \right]^2 \left( S^{\mathsf{T}} L(p) S \right)_{\sfa\sfa} \left[ \Pi_{\sfa}(p+q_{0}) - \Pi_{\sfa}(p) + \Pi_{\sfa}(p-q_{0}) - \Pi_{\sfa}(p) \right] \\
			\begin{split}
				&= \frac{e^2}{q_{0}^2} \frac{1}{\beta V} \sum_{p}  \sum_{\sfa} \left[ \gamma_{\sfa}^{i}(p,p) \right]^2 \left( S^{\mathsf{T}} L(p) S \right)_{\sfa\sfa} \Big[ \left(S^{\mathsf{T}} L^{-1}(p-q_{0}) S\right) - \left(S^{\mathsf{T}} L^{-1}(p) S\right) \\
				&\hspace{17.3em} + \left(S^{\mathsf{T}} L^{-1}(p+q_{0}) S\right) - \left(S^{\mathsf{T}} L^{-1}(p) S\right)\Big]_{\sfa\sfa}.
			\end{split}
		\end{align}
		The terms in the second line inside the square brackets can be brought into a similar form as the first line by shifting the momenta $p\mapsto p - q_{0}$
		\begin{align}
			\begin{split}
				K^{ii}_{\mathrm{MT}}(q_0,\v{0})  + K^{ii}_{\mathrm{DOS}}(q_0,\v{0}) &= \frac{e^2}{q_{0}^2} \frac{1}{\beta V} \sum_{p}  \sum_{\sfa}  \left[ \gamma_{\sfa}^{i}(p,p) \right]^2 \Big[ \left( S^{\mathsf{T}} L(p) S \right) - \left( S^{\mathsf{T}} L(p- q_{0}) S \right) \Big]_{\sfa\sfa} \\
				&\hspace{11em} \times \Big[ \left( S^{\mathsf{T}} L^{-1}(p-q_{0}) S \right) - \left( S^{\mathsf{T}} L^{-1}(p) S \right) \Big]_{\sfa\sfa}.
			\end{split}
		\end{align}
		To cast this into a form similar to the AL diagram, we use the fact that
		\begin{equation}
			\left( S^{\mathsf{T}} L(p) S \right)_{\sfa\sfa} - \left( S^{\mathsf{T}} L(p- q_{0}) S \right)_{\sfa\sfa} = \sum_{\sfb\sfc} \left( S^{\mathsf{T}} L(p) S \right)_{\sfa\sfb} \left[ \left( S^{\mathsf{T}} L^{-1}(p-q_{0}) S \right) - \left( S^{\mathsf{T}} L^{-1}(p) S \right)  \right]_{\sfb\sfc} \left( S^{\mathsf{T}} L(p-q_{0}) S \right)_{\sfc\sfa}.
		\end{equation}
		Using the fact that the difference of the inverse pair fluctuation propagators is diagonal in the band basis we find
			\begin{align}
			\begin{split}
				K^{ii}_{\mathrm{MT}}(q_0,\v{0})  + K^{ii}_{\mathrm{DOS}}(q_0,\v{0}) &= \frac{e^2}{q_{0}^2} \frac{1}{\beta V} \sum_{p} \sum_{\sfa\sfb}   \left[ \gamma_{\sfa}^{i}(p,p) \right]^2 \left( S^{\mathsf{T}} L(p) S \right)_{\sfb\sfa} \left( S^{\mathsf{T}} L(p-q_{0}) S \right)_{\sfa\sfb} \\
				& \times \Big[ \left( S^{\mathsf{T}} L^{-1}(p-q_{0}) S \right) - \left( S^{\mathsf{T}} L^{-1}(p) S \right) \Big]_{\sfb\sfb} \Big[ \left( S^{\mathsf{T}} L^{-1}(p-q_{0}) S \right) - \left( S^{\mathsf{T}} L^{-1}(p) S \right) \Big]_{\sfa\sfa}.
			\end{split} \\
			&= \frac{e^2}{\beta V }\sum_{p} \sum_{\sfa\sfb} \left[ \gamma^{i}_{\sfa}(p,p)\right]^2 C_{\sfa}(p,q_{0}) C_{\sfb}(p,q_{0}) \left( S^{\mathsf{T}} L(p) S \right)_{\sfb\sfa} \left( S^{\mathsf{T}} L(p-q_{0}) S \right)_{\sfa\sfb}. \label{eq:MT+DOS_uniform}
		\end{align}
		
	\subsection{Regular longitudinal conductivity}
		
		Adding the contribution in Eq.~\eqref{eq:MT+DOS_uniform} to the AL diagram in Eq.~\eqref{eq:AL_uniform_limit} yields
		\begin{equation}\label{eq:regular_conductivity_diagrams_uniform_limit}
			K^{ii}_{\mathrm{AL} + \mathrm{MT} + \mathrm{DOS}}(q_{0},\v{0}) = \frac{e^2}{\beta V }\sum_{p, \sfa\sfb } \gamma_{\sfa}^{i}(p,p) \left[ \gamma_{\sfa}^{i}(p,p) - \gamma_{\sfb}^{i}(p,p) \right] C_{\sfa}(p,q_{0}) C_{\sfb}(p,q_{0}) \left( S^{\mathsf{T}} L(p) S \right)_{\sfb\sfa} \left( S^{\mathsf{T}} L(p-q_{0}) S \right)_{\sfa\sfb}.
		\end{equation}
		Equation \eqref{eq:regular_conductivity_diagrams_uniform_limit} is a central result of this paper. 
        This equation is illustrated graphically in Fig.~\ref{fig:effective_bubbles}.
		\begin{figure*}[t]
			\centering
			\includegraphics[width=0.55\textwidth]{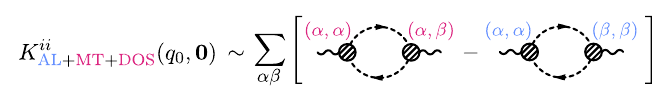}
			\caption{Schematic graphical representation of the fluctuation kernel that contributes to the regular conductivity in the uniform limit. 
				Here, the shaded circles represent the vertex block $ \gamma^{i}_{\sfa}(p,p) C_{\sfb}(p,q_{0})$ with the indices $(\sfa,\sfb)$.}
			\label{fig:effective_bubbles}
		\end{figure*}
	\end{widetext}
	
	At this stage, some remarks are in order. 
	First of all, it is a well-known fact that the MT and DOS diagrams cancel each other in the ultraclean limit when these are computed using a Fermi-surface average and assuming $T\tau$ finite but large \cite{Larkin-Varlamov-TheoryFluctuationsSuperconductors-2005,Reggiani-Varlamov-FluctuationConductivityLayered-1991,Livanov-Varlamov-StrongCompensationQuantum-2000,Stepanov-Skvortsov-SuperconductingFluctuationsArbitrary-2018}.
	Without making such approximations, we see here that such a cancellation is not exact.
	The leftover term was mapped to a form similar to the AL diagram, which brings out a new aspect of the theory of fluctuation conductivity in ultraclean metals.
	Namely, the conventional lore that the AL diagram should dominate over the MT and DOS diagrams close to the critical point because it is comprised of two exceedingly singular pair-fluctuation propagators rather than one is simply not correct. 
	An exact rewriting of the diagrams in this limit shows that they can all be brought into a form with two pair-fluctuation propagators and the same triangle block vertex.
	We emphasize that this delicate situation would not be realized if the interaction kernel were computed in a nonconserving approximation.
    Indeed, in most of the previous studies of the fluctuation conductivity of multicomponent superconductors only the AL contribution was computed \cite{Fanfarillo-Grilli-TheoryFluctuationConductivity-2009,Koshelev-Vinokur-TheoryFluctuationsTwoband-2005}.
    A justification for including only this diagram can be made only in the dirty limit.
    
	Moreover, this rewriting demonstrates that the multicomponent character of the incipient superconducting order and the parent metallic state plays a key role in determining the longitudinal conductivity.
	Specifically, when adding the contributions relevant for the regular part of the longitudinal conductivity the result in Eq.~\eqref{eq:regular_conductivity_diagrams_uniform_limit} is seen to vanish if (i) the effective masses of the bare electron bands are equal, or (ii) there is no interband coupling: $g_{\sfa\sfb} = 0$ for $\sfa\neq\sfb$.
	In other words, to obtain a finite regular conductivity, we need at least two Fermi surfaces with different Fermi velocities, each supporting the formation of Cooper pairs that can tunnel into each other.
    This implies that the cancellation is exact in the single-band case, a result that was also obtained in Ref.~\cite{Boyack-Boyack-RestoringGaugeInvariance-2018a}.
	
	That an ultraclean, translationally invariant system with only momentum-conserving electron-electron interactions can give rise to a finite dissipative part of the conductivity is somewhat counterintuitive. 
	Indeed, under these circumstances the usual reason for the nonrelaxation of the electrical current is conservation of momentum. 
	However, since the current vertex in the uniform limit involves the velocity operator and not the momentum, this symmetry protection fails in a multiband metal where the bands have different effective masses. 
	These systems therefore support an intrinsic finite regular conductivity.
	
	The ultraclean limit of the conventional fluctuation theory has been studied before \cite{Reggiani-Varlamov-FluctuationConductivityLayered-1991,Aronov-Larkin-GaugeInvarianceTransport-1995,Livanov-Varlamov-StrongCompensationQuantum-2000}.
	In these pioneering works, it was shown that the triangle block vertex $\Lambda^{\mu}(p-q_{0},p)$ (in effect, also $C_{\sfa}(p,q_{0})$) is a nonanalytic function of the external Matsubara frequency $q_{0}$, making it challenging to perform the analytical continuation to real frequencies.
	In fact, it was shown that $C_{\sfa}(p,q_{0} \neq 0 ) \equiv 0$, and a nonzero result was found only for $q_{0} = 0$.
	To proceed with the Matsubara summation of Eq.~\eqref{eq:regular_conductivity_diagrams_uniform_limit}, we therefore use this fact and assume that $q_{0} = 0$ in the triangle blocks $C_{\sfa}(p,q_{0})$, but carry on their $p_{0}$ dependence, so that some of the analytic structure of the block is included.
	This has the consequence that the block $C_{\sfa}(p,0)$ is exactly a derivative of the inverse pair-fluctuation propagator with respect to its Matsubara frequency.
    The physical meaning of this simplification is that the pole structure of the triangle block responsible for the indirect acceleration of Cooper pairs by the electric field is neglected \cite{Aronov-Larkin-GaugeInvarianceTransport-1995}.

	The Matsubara summation can then be converted to a complex contour integral in the standard fashion \cite{Altland-Simons-CondensedMatterField-2023,Larkin-Varlamov-TheoryFluctuationsSuperconductors-2005}.
	The calculation of this contour integral faces the same difficulties as in the conventional fluctuation theory (See Ref.~\cite{Larkin-Varlamov-TheoryFluctuationsSuperconductors-2005} for details).
	In particular, the pair-fluctuation propagator continued to the complex plane $L_{ab}(z, \v{p})$ has a branch cut along the real axis, meaning that the integrand of Eq.~\eqref{eq:regular_conductivity_diagrams_uniform_limit} will have branch cuts along $\Im(z)=0$ and $\Im(z)=\Omega_m$, where $\iu \Omega_m = q_{0}$.
    Similarly, the triangle block $C_{\sfa}(z,\v{p})$ also has a cut along $\Im(z)=0$.
	Deforming the contour around these cuts results in (See Refs.~\cite{Larkin-Varlamov-TheoryFluctuationsSuperconductors-2005,Reggiani-Varlamov-FluctuationConductivityLayered-1991})
	\begin{widetext}
		\begin{align}\label{eq:K_fluct_reg_uniform}
				K^{ii}_{\mathrm{AL} + \mathrm{MT} + \mathrm{DOS}}(q_{0},\v{0}) &= e^2 \int \frac{\D^d p}{(2\pi)^d} \int_{-\infty}^{\infty}\frac{\D z}{2\pi} \coth\left(\frac{z}{2T}\right) \sum_{\sfa\sfb} \gamma_{\sfa}^{i}(\v{p},\v{p}) \left[ \gamma_{\sfa}^{i}(\v{p},\v{p}) - \gamma_{\sfb}^{i}(\v{p},\v{p}) \right] \notag \\
				&\times \bigg[ \Im\left\{C_{\sfa}^{R}(z,\v{p}) C_{\sfb}^{R}(z,\v{p}) \left( S^{\mathsf{T}} L^{R}(z,\v{p}) S \right)_{\sfa\sfb} \right\}  \left\{ \left( S^{\mathsf{T}} L^{R}(z+\iu\Omega_m,\v{p}) S \right) + \left( S^{\mathsf{T}} L^{A}(z-\iu\Omega_m,\v{p}) S \right) \right\}_{\sfb\sfa} \notag  \\
				&\hspace{0.75em}+ \Im\left\{ C_{\sfa}^{A}(z,\v{p}) C_{\sfb}^{A}(z,\v{p}) \right\} \left( S^{\mathsf{T}} L^{R}(z,\v{p}) S \right)_{\sfa\sfb} \left( S^{\mathsf{T}} L^{A}(z-\iu\Omega_m,\v{p}) S \right)_{\sfb\sfa} \bigg].
		\end{align}
	\end{widetext}
	Here, we have introduced the notation $L^{R/A}(z)$ for the analytic function above (R) and below (A) the branch cut at $\Im(z) = 0$. 
    To arrive at this equation, we have reused many arguments of Ref.~\cite{Reggiani-Varlamov-FluctuationConductivityLayered-1991}.
    Notably, we neglect the frequency dependence of the triangle blocks acquired from the shift of integration variables in the integrals around the branch cut at $\Im(z) = \Omega_m$. 
    At this point, the analytical continuation $q_{0} = \iu\Omega_m \to \omega + \iu 0$ can safely be performed.
	
	\subsection{Explicit calculation in a two-band model}\label{sec:twoband}
	
	As a concrete example, we now compute the static limit of the regular conductivity in a two-band model.
	We choose an interaction matrix $\mathbb{G} = (g_{\sfa\sfb})$ given by 
	\begin{equation}
		\mathbb{G} = \begin{pmatrix}
			g_1 & \delta g \\
			\delta g & g_2 
		\end{pmatrix}
		= g \begin{pmatrix}
			1 + \kappa & \eta \\
			\eta & 1 - \kappa
		\end{pmatrix},
	\end{equation}
	where $0 < \sqrt{\eta^2 + \kappa^2} < 1$.
	The inverse pair-fluctuation propagator in the band basis is then
	\begin{align}\label{eq:L_inv_twoband}
		&(S^{\mathsf{T}} L^{-1}(q) S)_{\sfa\sfb} = \Pi_{\sfa\sfb}(q) - (\mathbb{G}^{-1})_{\sfa\sfb} \notag \\
		&= \frac{1}{\tilde{g}} \begin{pmatrix}
			\tilde{g} \Pi_{1}(q) - (1-\kappa) & \eta \\
			\eta & \tilde{g} \Pi_{2}(q) - (1+\kappa)
		\end{pmatrix},
	\end{align}
	where $\tilde{g} \coloneqq g (1 - \eta^2 - \kappa^2)$.
	The mean-field critical temperature is determined by the \textit{largest} temperature solving the equation $\det L^{-1}(0,\v{0}) = 0$ \cite{Suhl-Walker-BardeenCooperSchriefferTheorySuperconductivity-1959,Koshelev-Vinokur-TheoryFluctuationsTwoband-2005}.
    The frequency and momentum dependence of the pair-fluctuation propagator is inherited from the particle-particle bubble (See Eq.~\eqref{eq:Pi_large_freq} in Appendix \ref{app:fluctuation_prop}).

    \begin{figure}[htb]
		\centering
		\includegraphics[width=\columnwidth]{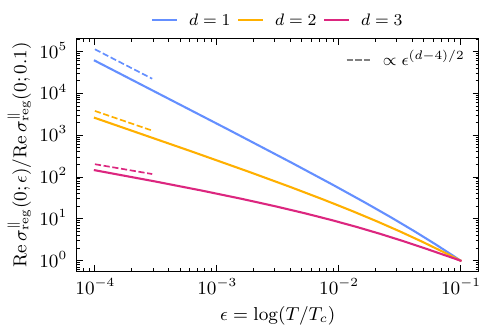}
		\caption{Regular part of longitudinal conductivity at $\omega = 0$ as a function of reduced temperature $\epsilon$, normalized with respect to its value at $\epsilon =0.1$. The parameters used here are $m_2 = 2 m_1$, $\kappa = 0.7$ and $\eta = 0.2$.
        }
		\label{fig:reg_sigma}
	\end{figure}

    Numerical calculation of the regular conductivity from Eq.~\eqref{eq:K_fluct_reg_uniform} in the static limit $\omega\to 0$ yields a non-vanishing result only if the effective masses are different and $\eta\neq0$, in agreement with the selection rules identified before.
    The temperature dependence of this conductivity is shown in Fig.~\ref{fig:reg_sigma}.
    Close to the critical temperature, the conductivity receives a logarithmically divergent contribution in $T/T_c$ due to the Gaussian-level fluctuation corrections. 
    Its exponent is approximately given by $(d-4)/2$, which coincides with the classical results in the diffusive limit \cite{Aslamazov-Larkin-EffectFluctuationsProperties-1968,Larkin-Varlamov-TheoryFluctuationsSuperconductors-2005}.
    However, we emphasize that, in the present case, this is a purely intrinsic contribution to the conductivity.
    The details in the prefactor of this correction are of secondary importance, insofar as it is nonzero, and therefore they are not presented here.

    The conjectured divergence illustrated with dashed lines in Fig.~\ref{fig:reg_sigma} can be obtained by approximating the vertex blocks $C_{\sfa}^{R}(z,\v{p})$ in Eq.~\eqref{eq:K_fluct_reg_uniform} by their static value $C_{\sfa}^{R}(0,\v{p})$.
    Combined with Eq.~\eqref{eq:sigma_reg_def}, a Taylor expansion of the analytic continuation of Eq.~\eqref{eq:K_fluct_reg_uniform} to leading order in $\omega$ now yields
        \begin{align}\label{eq:DC_regular}
                &\Re\sigma_{\mathrm{reg}}^{\parallel}(0) \simeq - \frac{e^2}{2T} \int \frac{\D^d p}{(2\pi)^d} \int_{-\infty}^{\infty}\frac{\D z}{2\pi} \frac{1}{\sinh^2(z/2T)} \notag \\
                &\quad\times \sum_{\sfa\sfb} \gamma_{\sfa}^{i}(\v{p},\v{p}) \left[ \gamma_{\sfa}^{i}(\v{p},\v{p}) - \gamma_{\sfb}^{i}(\v{p},\v{p}) \right] \\
            &\quad\times C_{\sfa}^{R}(0,\v{p}) C_{\sfb}^{R}(0,\v{p}) \left[ \Im  \left( S^{\mathsf{T}} L^{R}(z,\v{p}) S \right)_{\sfa\sfb} \right]^2. \notag
        \end{align}
    Here we have used the fact that $(S^{\mathsf{T}} L(p) S)$ is symmetric and $C_{\sfa}^{R}(0,\v{p}) C_{\sfa}^{R}(0,\v{p})$ is real.
    After inserting the expressions for the vertices and the pair-fluctuation propagator, one finds that the integrand has a similar functional dependence of $z$ and $p$ to that in the classical AL calculation \cite{Larkin-Varlamov-TheoryFluctuationsSuperconductors-2005}.
    This implies the same temperature dependence as the AL paraconductivity as the critical temperature is approached, i.e., $\Re \sigma^{\parallel}_{\mathrm{reg}}(0) \sim \epsilon^{(d-4)/2}$ \cite{Aslamazov-Larkin-InfluenceFluctuationPairing-1968,Aslamazov-Larkin-EffectFluctuationsProperties-1968}.
    However, note that only the off-diagonal components of $(S^{\mathsf{T}} L(p) S)$ can contribute to the sum.
    For the two-band example, this implies that $\Re \sigma^{\parallel}_{\mathrm{reg}}(0) \propto \eta^2  $.

    Although the case of an $(N\geq3)$-component superconductor is principally different from the two-component case considered above, Eq.~\eqref{eq:DC_regular} shows that its fluctuation conductivity in the normal state is governed by the same selection rules as the two-component case.
    One of the features that separates $N=2$ and $N\geq 3$ is the possibility of a mean-field state with broken time-reversal symmetry.
    Since this is known to occur \textit{below} $T_c$ \cite{Stanev-Tesanovic-ThreebandSuperconductivityOrder-2010}, it is expected not to affect the fluctuation response in the normal state.
    However, recent experiments show hints of time-reversal symmetry breaking extending into the normal state of the metal \cite{Grinenko-Babaev-StateSpontaneouslyBroken-2021,Shipulin-Grinenko-CalorimetricEvidenceTwo-2023}.
    A proper account of its transport properties requires going beyond the Gaussian fluctuation approximation \cite{Samoilenka-Babaev-MicroscopicTheoryElectron-2026}, and is therefore not accessible in the present theory.

    
	\subsection{Frequency $f$-sum rule}\label{sec:sum_rule}
	
	For completeness, let us also address the effects of the DIA diagram, as was done in the single-component case by Ref.~\cite{Boyack-Boyack-RestoringGaugeInvariance-2018}.
	This analysis is a multicomponent generalization of these results.
	
	Being purely real and independent of $q$, $K^{\mu\nu}_{\mathrm{DIA}}(q)$ only affects the Drude weight of the conductivity tensor.
	Introducing temporarily a $q_{0}$-dependence by using $G_{0,\sfa}(k) G_{0,\sfa}(k) = \lim_{q_{0}\to 0} G_{0,\sfa}(k) G_{0,\sfa}(k-q_{0})$ together with the identity in Eq.~\eqref{eq:q_0_identity} yields
    \begin{widetext}
		\begin{align}
			K^{\mu\nu}_{\mathrm{DIA}}(q_{0},\v{0}) &= \lim_{q_{0}\to 0}\sum_{\sfa} \frac{2e^2}{m_{\sfa}} \frac{1}{(\beta V)^2} \delta^{\mu\nu}(1 - \delta^{\mu 0}) \sum_{pk} \left(S^{\mathsf{T}} L(p) S\right)_{\sfa\sfa} \frac{1}{q_{0}} \left[ G_{0,\sfa}(k-q_{0}) - G_{0,\sfa}(k) \right]  G_{0,\sfa}(p-k) \notag \\
			&=\lim_{q_{0}\to 0}\sum_{\sfa} \frac{2e^2}{m_{\sfa}} \delta^{\mu\nu}(1 - \delta^{\mu 0}) \frac{1}{\beta V}  \sum_{p} \left(S^{\mathsf{T}} L(p) S\right)_{\sfa\sfa}  \frac{1}{q_{0}} \left[ \Pi_{\sfa}(p-q_{0}) - \Pi_{\sfa}(p) \right] \notag \\
			&= -\sum_{\sfa} \frac{2e^2}{m_{\sfa}} \delta^{\mu\nu}(1 - \delta^{\mu 0}) \frac{1}{\beta V} \sum_{p} \left(S^{\mathsf{T}} L(p) S\right)_{\sfa\sfa} \frac{\partial}{\partial p_{0}} \left(S^{\mathsf{T}} L^{-1}(p) S\right)_{\sfa\sfa} .
		\end{align}
	\end{widetext}
	After using that $2 \times \partial L^{-1}(p)/\partial p_{0} = \partial L^{-1}(p) /\partial \mu$ we can write the contribution to the Drude weight as 
	\begin{equation}
		K^{ii}_{\mathrm{DIA}}(q_{0},\v{0}) = \sum_{\sfa} \frac{e^2 n_{\mathrm{fluc},\sfa} }{m_{\sfa}},
	\end{equation}
	where the fluctuation correction to the number density of flavor $\sfa$ is given by \cite{Boyack-Boyack-RestoringGaugeInvariance-2018}
	\begin{equation}
		\sum_{\sfa } n_{\mathrm{fluc},\sfa} \equiv - \frac{1}{\beta V} \frac{\partial}{\partial \mu} \tr \log\left(-\beta L^{-1}\right).
	\end{equation}
	As shown by Boyack in Ref.~\cite{Boyack-Boyack-RestoringGaugeInvariance-2018}, the DIA diagram exhausts the right-hand side of the conductivity $f$-sum rule
	\begin{equation}
		\int_{-\infty}^{+\infty} \frac{\D\omega}{2\pi} \Re \sigma^{\parallel}(\omega) = \sum_{\sfa} \frac{n_{\sfa} e^2}{2m_{\sfa}}.
	\end{equation}
	To satisfy the sum rule, we must therefore make sure the real part of the residual response kernel $\delta K^{ii}(\omega+\iu0) \coloneqq K_{\mathrm{fluc}}^{ii}(\omega+\iu0) - K_{\mathrm{DIA}}^{ii}(\omega+\iu0)$ multiplied by $\pi$ and evaluated at $\omega \to 0$ cancels exactly the integral of the regular conductivity.
	After some rearrangements, one finds that this is identically ensured by the Kramers-Kronig relations for the function $\delta K^{ii}(\omega + \iu0)$.
	This is yet another demonstration that the Gaussian fluctuation approximation is conserving.
	
	\section{Conclusion}\label{sec:conc}

    In this paper, we have presented a detailed derivation of the electromagnetic response tensor due to fluctuating Cooper pairs above $T_c$ in a multicomponent superconductor. 
    By using the formalism developed in Ref.~\cite{Boyack-Boyack-RestoringGaugeInvariance-2018}, we showed that the Gaussian-fluctuation approximation is conserving in the sense that it is not conflicting with local charge conservation.
    This level of consistency is required in any theory of transport in a many-body system. 
    
    Within the same formalism, we also derived the fluctuation corrections to the DC conductivity close to the superconducting critical point.
    Ignoring disorder and impurities from the outset, we find that the common separation in terms of the Aslamazov-Larkin (AL), Maki-Thompson (MT) and density-of-states (DOS) diagrams is somewhat artificial since they partially cancel one another. 
    Since the pair-fluctuation propagator has a pole at the critical point, usually, only the AL diagram (containing two such propagators) is considered in this regime. 
    In the ultraclean limit, this rationale fails and the partial cancellation with the remaining diagrams is only revealed after including the same diagrams that produce a consistent approximation of the linear response theory.
    A similar conclusion was recently made in a study of the fluctuation theory using the Keldysh formalism \cite{Stepanov-Skvortsov-SuperconductingFluctuationsArbitrary-2018}.
    
    The ultraclean multicomponent superconductor has a correction to the dissipative part of the conductivity with no counterpart in single-band systems. 
    Indeed, the complete set of fluctuation diagrams produces a term logarithmically divergent in $T/T_c$ only if there are at least two electron bands that (i) have different effective masses and (ii) can take part in interband scattering of Cooper pairs.
    These conditions cannot be met in a single-band superconductor.
    Previous studies of the fluctuation conductivity in multicomponent superconductors did not identify this mechanism. This can be attributed to the fact that they were concerned with the dirty limit \cite{Dzero-Levchenko-TransportAnomaliesMultiband-2023} or considered only the AL diagram \cite{Koshelev-Vinokur-TheoryFluctuationsTwoband-2005}.
    The case of dominant interband pairing in a multiband superconductor was considered in Ref.~\cite{Fanfarillo-Grilli-TheoryFluctuationConductivity-2009}.
    They concluded that the fluctuation conductivity takes the same form as in the single-band case, but did not discuss the effect of the MT and DOS diagrams. 
    A direct comparison with the present derivation is, however, complicated by the fact that dominant interband pairing contradicts our assumption that the interaction matrix $\mathbb{G} = (g_{\sfa\sfb})$ be positive definite (See Eq.~\eqref{eq:hamiltonian}).

    The divergence of this correction should not be confused with the Drude term of the conductivity tensor, resulting from the absence of impurities. 
    The logarithmically divergent fluctuation correction is a separate contribution that derives from the proximity to the superconducting transition, liberating an increasing amount of fluctuating pairs.
    The absence of this conductivity correction in a single-band system can be attributed to translational symmetry, which enforces non-relaxation of the current.
    In a multi-band system where the electrons have different effective masses, this symmetry protection fails because the current is related to the velocity operator rather than the momentum operator itself. 
    A similar mechanism enabling a finite optical conductivity in a clean, translationally invariant system was recently considered in Ref.~\cite{Gindikin-Maslov-QuantumCriticalityOptical-2024}, where it was dubbed an ``intervalley drag". 
    In the present case, this mechanism is only revealed in combination with a physically meaningful (i.e., gauge-invariant) theory of superconducting fluctuations. This connection shows that transport in multicomponent superconductors is more subtle than previously appreciated.
    
	\begin{acknowledgements}
		We thank Erlend Syljuåsen and Yuriy Yerin for useful discussions,
        and Rufus Boyack for pointing out Ref.~\cite{Boyack-Boyack-RestoringGaugeInvariance-2018a}.
		We acknowledge support from the Norwegian Research Council through Grant No. 262633, ``Center of Excellence on Quantum Spintronics” and Grant No. 323766, as well as COST Action CA21144 ``Superconducting Nanodevices and Quantum Materials for Coherent Manipulation".
	\end{acknowledgements}
	
	\appendix

	\onecolumngrid
	
	\section{Pair-fluctuation propagator}\label{app:fluctuation_prop}
	
	In this appendix, we provide some details on the computation of the pair fluctuation propagator
	\begin{equation}
		\left[ - L^{-1}(x-y) \right]_{ab} = \frac{\delta_{ab}}{\lambda_{a}} \delta(x-y) - \frac{\delta^2}{\delta \bar{\phi}_{a}(x) \delta \phi_{b}(y) } \tr \log \left(-\beta \mathcal{G}^{-1}[\bar{\phi},\phi]\right)\biggr\lvert_{\bar{\phi},\phi = 0}.
	\end{equation}
	Owing to the saddle-point condition $\delta S_{\mathrm{HS}}[\bar{\phi},\phi]/\delta \phi = 0$, there are no terms linear in $\eta$ in $S_{\mathrm{eff}}[\bar{\eta},\eta]$.
	Performing the two functional derivatives and resolving the functional trace yields
	\begin{equation}
		\left[ - L^{-1}(x-y) \right]_{ab} = \frac{\delta_{ab}}{\lambda_{a}} \delta(x-y) - \int \D z \int \D z' \tr \left( \frac{\delta \mathcal{G}(z,z')}{\delta \bar{\phi}_{a}(x)} \frac{\delta \mathcal{G}^{-1}(z',z)}{\delta \phi_{b}(y)} \right),
	\end{equation}
	where the remaining trace is a matrix trace.
	Using the identity $\int \D x' \mathcal{G}(x,x') \mathcal{G}^{-1}(x',y) = \delta(x-y)$ we can write
	\begin{equation}\label{eq:delta_G}
		\delta \mathcal{G}(z,z') = - \int \D y \int \D y' \mathcal{G}(z,y) \delta \mathcal{G}^{-1}(y,y') \mathcal{G}(y',z').
	\end{equation} 
	This yields
	\begin{equation}
		\left[- L^{-1}(x-y) \right]_{ab} =  \frac{\delta_{ab} \delta(x-y)}{\lambda_a} 
		+ \int \prod_{i=1}^{4} \D z_{i} \tr \left( \mathcal{G}(z_1,z_2) \frac{\delta \mathcal{G}^{-1}(z_{2},z_{3})}{\delta \bar{\phi}_{a}(x)} \mathcal{G}(z_{3},z_4) \frac{\delta \mathcal{G}^{-1}(z_{4},z_{1})}{\delta \phi_{b}(y)}  \right). 
	\end{equation}
	The functional derivatives of the inverse Green's function in Eq.~\eqref{eq:Greens-function} are easily computed and this finally yields
	\begin{equation}
		\left[- L^{-1}(x-y) \right]_{ab} = \frac{\delta_{ab} \delta(x-y)}{\lambda_a} + \sum_{\sfa} s_{a\sfa} s_{b\sfa} G_{0,\sfa}(x,y) \tilde{G}_{0,\sfa}(y,x).
	\end{equation}
	
	To write the Gaussian fluctuation action in the Fourier basis, we introduce the Fourier representation of the fields as
	\begin{equation}
		\eta_{a}(x) = \frac{1}{\sqrt{\beta V}} \sum_{q} \e^{\iu q \cdot x} \eta_{a}(q) \quad \text{and} \quad \eta_{a}(q) = \frac{1}{\sqrt{\beta V}} \int \D x \, \e^{-\iu q \cdot x} \eta_{a}(x).
	\end{equation}
	Here we have defined the four-momentum as $q^{\mu} \equiv (q_0,\v{q}) = (\iu\Omega_{m}, \v{q})$, where $\Omega_m$ is a bosonic Matsubara frequency and the summation is $\sum_{q} \equiv \sum_{m\in \ZZ} \sum_{\v{q}}$. 
	Moreover, note that $q \cdot x \coloneqq \v{q} \cdot \v{r} - \Omega_{m} \tau \equiv -\mathfrak{g}_{\mu\nu} x^{\mu} q^{\nu}$.
	For fermionic fields, we employ the same convention with fermionic Matsubara frequencies instead of bosonic ones.
	
	Using these definitions, the effective Gaussian fluctuation action reads
	\begin{align}
		S_{\mathrm{eff}}[\bar{\eta},\eta] &= - \tr\log(-\beta \mathcal{G}^{-1}_{0}) + \int \D x \int \D y \sum_{ab} \bar{\eta}_{a}(x) \left[- L^{-1}(x-y) \right]_{ab} \eta_{b}(y) \notag \\
		&=  - \tr\log(-\beta \mathcal{G}^{-1}_{0}) + \sum_{q} \sum_{ab} \bar{\eta}_{a}(q) \left[ - L^{-1}(q) \right]_{ab} \eta_{b}(q),
	\end{align}
		with
	\begin{align}\label{eq:L_inv_explicit}
		- L^{-1}_{ab}(q) &= \int \D x\, \e^{-\iu q \cdot x} \left[ - L^{-1}(x) \right]_{ab} = \frac{\delta_{ab}}{\lambda_a} + \sum_{\sfa} s_{a\sfa} s_{b\sfa} \frac{1}{\beta V} \sum_{k} G_{0,\sfa}(k) \tilde{G}_{0,\sfa}(k-q).
	\end{align}
	Here, the Fourier representation of the fermion Green's functions is analogously introduced as
	\begin{equation}\label{eq:Fourier_transformed_G}
		G_{0,\sfa}(x,y) = \frac{1}{\beta V} \sum_{k} \e^{\iu k \cdot(x-y)} G_{0,\sfa}(k). 
	\end{equation}
	After using that $\tilde{G}_{0,\sfa}(k) = -G_{0,\sfa}(-k)$ we find
	\begin{equation}
		L^{-1}_{ab}(q) = - \frac{\delta_{ab}}{\lambda_a} + \left( S \Pi(q) S^{\mathsf{T}} \right)_{ab},
	\end{equation}
	where $\Pi_{\sfa\sfb}(q)$ denotes the particle-particle bubble, as defined in Eq.~\eqref{eq:Pi}.
	
	Let us consider the particle-particle bubble more closely
	\begin{equation}
		\Pi_{\sfa}(\iu\Omega_m,\v{q}) = T \sum_{n\in\ZZ} \int \frac{\D^d p}{(2\pi)^d} G_{0,\sfa}(\iu\omega_n - \iu\Omega_m,\v{p}-\v{q}) G_{0,\sfa}(-\iu\omega_n-\v{p}).
	\end{equation}
	In the long-wavelength approximation we linearize the spectrum about the Fermi surface $\xi_{\sfa}(\v{p}-\v{q}) \simeq \xi_{\sfa}(\v{p})- \v{q} \cdot \v{v}_{\mathrm{F},\sfa}$, and we convert the momentum integral to an energy integral by using 
	\begin{equation}
		\int \frac{\D^d p}{(2\pi)^d} F(\xi_{\sfa}(\v{p})) \simeq \nu_{\sfa} \int \D \xi F(\xi),
	\end{equation}
    where $\nu_{\sfa}$ denotes the density of states at the Fermi level of flavor $\sfa$.
	Performing the integral over $\xi$ in the particle-particle bubble can now easily be done with the help of the Cauchy integral theorem. 
	The remaining Matsubara summation is formally divergent, but is regularized by introducing the Debye frequency $\omega_{D}$ as an upper cutoff.
	The resulting long-wavelength particle-particle bubble reads \cite{Larkin-Varlamov-TheoryFluctuationsSuperconductors-2005}
	\begin{equation}
			\Pi_{\sfa}(\iu\Omega_m, \v{q}) \simeq \nu_{\sfa} \bigg[ \log\left(\frac{2 \omega_D \e^{\gamma_{E}}}{\pi T}\right) - \uppsi\left(\frac{1}{2} + \frac{\abs{\Omega_m}}{4 \pi T}\right) + \uppsi\left(\frac{1}{2}\right) + \frac{v_{\mathrm{F},\sfa}^2}{32 d \pi^2} q^2  \uppsi''\left(\frac{1}{2} + \frac{\abs{\Omega_m}}{4 \pi T}\right) \bigg],		
	\end{equation}
	where $\uppsi(x)$ denotes the digamma function.
	Assuming that the critical temperature takes the BCS form 
	\begin{equation}
		T_c = \frac{2 \omega_D \e^{\gamma_{\mathrm{E}}}}{\pi} \e^{-x},
	\end{equation}
	we can write 
	\begin{equation}
			\Pi_{\sfa}(\iu\Omega_m, \v{q}) \simeq - \nu_{\sfa} \bigg[ \epsilon - x + \uppsi\left(\frac{1}{2} + \frac{\abs{\Omega_m}}{4 \pi T}\right) - \uppsi\left(\frac{1}{2}\right) - \frac{v_{\mathrm{F},\sfa}^2}{32 d \pi^2} q^2  \uppsi''\left(\frac{1}{2} + \frac{\abs{\Omega_m}}{4 \pi T}\right) \bigg],\label{eq:Pi_large_freq}
	\end{equation}
	where $\epsilon \coloneqq \log(T/T_c)$.
    For the two-band model considered in the main text, we find from solving the equation $\det L^{-1}(0,\v{0})$ with Eq.~\eqref{eq:L_inv_twoband} that $x=x_{-}$, where
    \begin{equation}
			x_{\pm} = \frac{1}{2 \nu_{1} \nu_{2} \tilde{g}} \bigg[ \left( \nu_{1}(1+\kappa) + \nu_2 (1 - \kappa)   \right) \pm \sqrt{ \left( \nu_{1}(1+\kappa) + \nu_2 (1 - \kappa)   \right)^2 - 4 \nu_1 \nu_2 \left( 1 - (\eta^2 + \kappa^2) \right) }\bigg].
\end{equation}
	
	\section{Fluctuation response kernel}\label{app:fluctuation_response}
	
	The fluctuation response kernel is obtained by performing two functional derivatives of the Gaussian fluctuation action $S_{\mathrm{fluc}}[A] = \tr\log(-\beta L^{-1}[A])$ with respect to $A^{\mu}$.
	The $A^{\mu}$-dependence of $L^{-1}$ is obtained by performing the minimal substitution $\partial_{\tau} \mapsto \partial_{\tau} - \iu e A_{0}$ and $-\iu\bm{\nabla} \mapsto - \iu\bm{\nabla} + e \v{A}$ in the bare particle Green's functions in Eq.~\eqref{eq:L_inv_explicit} and the same substitution with $e \to -e $ in the hole Green's functions.
	
	Let us first derive the expression for the three-point vertex in Eq.~\eqref{eq:threept_vertex}.
	Performing one functional derivative of the fluctuation action yields
	\begin{align}
		\frac{\delta L^{-1}_{ab}[A](y_1,y_2)}{\delta A_{\mu}(x)} &= -\frac{\delta}{\delta A_{\mu}(x)} \sum_{\sfa} s_{a\sfa} s_{b\sfa} G_{0,\sfa}[A](y_1,y_2) \tilde{G}_{0,\sfa}[A](y_2,y_1) \notag \\
		&= - \sum_{\sfa} s_{a\sfa} s_{b\sfa} \bigg[ \frac{\delta G_{0,\sfa}[A](y_1,y_2)}{\delta A_{\mu}(x)} \tilde{G}_{0,\sfa}[A](y_2,y_1) + G_{0,\sfa}[A](y_1,y_2) \frac{\delta \tilde{G}_{0,\sfa}[A](y_2,y_1)}{\delta A_{\mu}(x)} \bigg].
	\end{align}
	Using the property in Eq.~\eqref{eq:delta_G} leads to 
	\begin{equation}
		\begin{split}
			\frac{\delta L^{-1}_{ab}[A](y_1,y_2)}{\delta A_{\mu}(x)} &= \sum_{\sfa} s_{a\sfa} s_{b\sfa} \int \prod_{i=1}^{2} \D z_{i} G_{0,\sfa}[A](y_1,z_1) \frac{\delta G_{0,\sfa}^{-1}[A](z_1,z_2)}{\delta A_{\mu}(x)} G_{0,\sfa}[A](z_2,y_2) \tilde{G}_{0,\sfa}[A](y_2,y_1) \\
			&+\sum_{\sfa} s_{a\sfa} s_{b\sfa} \int \prod_{i=1}^{2} \D z_{i} G_{0,\sfa}[A](y_1,y_2) \tilde{G}_{0,\sfa}[A](y_2,z_1) \frac{\delta \tilde{G}_{0,\sfa}^{-1}[A](z_1,z_2)}{\delta A_{\mu}(x)} \tilde{G}_{0,\sfa}[A](z_2,y_1). \label{eq:L_one_derivative}
		\end{split}
	\end{equation}
	The two contributions above give equal contributions to the AL diagram in the end. 
	Indeed, if one makes a graphical representation of this vertex as in Fig.~\ref{fig:AL_vertex}, one finds two triangle blocks with opposite orientations.
	Setting $A = 0$ in Eq.~\eqref{eq:L_one_derivative} yields the vertex $\Lambda^{\mu}_{ab}(y_1,y_2;x)$ of Eq.~\eqref{eq:threept_vertex}
	\begin{equation}\label{eq:Lambda_mu}
		\Lambda^{\mu}_{ab}(y_1,y_2;x) = 2e\sum_{\sfa} s_{a\sfa} s_{b\sfa} \int \prod_{i=1}^{2} \D z_{i} G_{0,\sfa}(y_1,z_1) \gamma^{\mu}_{\sfa}(z_1,z_2;x) G_{0,\sfa}(z_2,y_2) \tilde{G}_{0,\sfa}(y_2,y_1),
	\end{equation}
	where the bare particle and hole vertices are given by
	\begin{equation}\label{eq:bare_particle_and_hole_vertex}
		e\gamma^{\mu}_{\sfa}(z_1,z_2;x) \coloneqq \frac{\delta G_{0,\sfa}^{-1}[A](z_1,z_2)}{\delta A_{\mu}(x)} \biggr\lvert_{A=0} \quad \text{and} \quad e\tilde{\gamma}^{\mu}_{\sfa}(z_1,z_2;x) \coloneqq \frac{\delta \tilde{G}_{0,\sfa}^{-1}[A](z_1,z_2)}{\delta A_{\mu}(x)} \biggr\lvert_{A=0}.
	\end{equation}
	
	The four-point vertex in Eq.~\eqref{eq:fourpt_vertex} is found by performing another functional derivative of Eq.~\eqref{eq:L_one_derivative} with respect to $A_{\nu}(x')$ and subsequently setting $A=0$.
	Together with repeated application of Eq.~\eqref{eq:delta_G} this yields
	\begin{align}\label{eq:L_two_derivatives}
		\Gamma^{\mu\nu}_{ab}(y_1,y_2;x,x') \hspace{5em} & \notag \\ 
		= - 2 \sum_{\sfa} s_{a\sfa} s_{b\sfa}\int \prod_{j} \D z_j \bigg[ &G_{0,\sfa}[A](y_1,z_1) \frac{\delta G_{0,\sfa}^{-1}[A](z_1,z_2)}{\delta A_{\mu}(x)} G_{0,\sfa}[A](z_2,z_3) \frac{\delta G_{0,\sfa}^{-1}[A](z_3,z_4)}{\delta A_{\nu}(x')} G_{0,\sfa}[A](z_4,y_2) \tilde{G}_{0,\sfa}[A](y_2,y_1)  \notag \\
		+\, &G_{0,\sfa}[A](y_1,z_1) \frac{\delta G_{0,\sfa}^{-1}[A](z_1,z_2)}{\delta A_{\nu}(x')} G_{0,\sfa}[A](z_2,z_3) \frac{\delta G_{0,\sfa}^{-1}[A](z_3,z_4)}{\delta A_{\mu}(x)} G_{0,\sfa}[A](z_4,y_2) \tilde{G}_{0,\sfa}[A](y_2,y_1) \notag \\
		+\, &G_{0,\sfa}[A](y_1,z_1) \frac{\delta G_{0,\sfa}^{-1}[A](z_1,z_2)}{\delta A_{\nu}(x')} G_{0,\sfa}[A](z_2,y_2) \tilde{G}_{0,\sfa}[A](y_2,z_3) \frac{\delta \tilde{G}_{0,\sfa}^{-1}[A](z_3,z_4)}{\delta A_{\mu}(x)} \tilde{G}_{0,\sfa}[A](z_4,y_1) \notag \\
		-\, &G_{0,\sfa}[A](y_1,z_1) \frac{\delta^2 G_{0,\sfa}^{-1}[A](z_1,z_2)}{\delta A_{\mu}(x) \delta A_{\nu}(x')} G_{0,\sfa}[A](z_2,y_2) \tilde{G}_{0,\sfa}[A](y_2,y_1) \bigg]\biggr\lvert_{A=0}.
	\end{align}
	Finally putting $A=0$ yields
	\begin{align}\label{eq:Gamma_munu}
		\Gamma^{\mu\nu}_{ab}(y_1,y_2;x,x') \hspace{9em} & \notag \\
		 = - 2e^2 \sum_{\sfa} s_{a\sfa} s_{b\sfa}\int \prod_{j} \D z_j \bigg[ &G_{0,\sfa}(y_1,z_1) \gamma^{\mu}_{\sfa}(z_1,z_2;x) G_{0,\sfa}(z_2,z_3) \gamma^{\nu}_{\sfa}(z_3,z_4;x') G_{0,\sfa}(z_4,y_2) \tilde{G}_{0,\sfa}(y_2,y_1) \notag \\
		+\, &G_{0,\sfa}(y_1,z_1) \gamma^{\nu}_{\sfa}(z_1,z_2;x') G_{0,\sfa}(z_2,z_3) \gamma^{\mu}_{\sfa}(z_3,z_4;x) G_{0,\sfa}(z_4,y_2) \tilde{G}_{0,\sfa}(y_2,y_1) \phantom{\bigg]} \notag \\
		+\, &G_{0,\sfa}(y_1,z_1) \gamma^{\nu}_{\sfa}(z_1,z_2;x') G_{0,\sfa}(z_2,y_2) \tilde{G}_{0,\sfa}(y_2,z_3) \tilde{\gamma}^{\mu}_{\sfa}(z_3,z_4;x) \tilde{G}_{0,\sfa}(z_4,y_1) \phantom{\bigg]} \notag \\
		-\, &G_{0,\sfa}(y_1,z_1) \gamma^{\mu\nu}_{\sfa}(z_1,z_2;x,x')G_{0,\sfa}(z_2,y_2) \tilde{G}_{0,\sfa}(y_2,y_1) \bigg],
	\end{align}
	where we have defined
	\begin{equation}\label{eq:bare_diamagnetic_vertex}
		e^2\gamma^{\mu\nu}_{\sfa}(z_1,z_2;x,x') \coloneqq \frac{\delta^2 G_{0,\sfa}^{-1}[A](z_1,z_2)}{\delta A_{\mu}(x) \delta A_{\nu}(x')}  \biggr\lvert_{A=0}.
	\end{equation}
	The first and second line of Eq.~\eqref{eq:Gamma_munu} yield the DOS diagrams, shown graphically in Fig.~\ref{fig:fluctuation_diagrams} \textbf{(b)} and \textbf{(c)}.
	The third line yields the MT diagram, shown graphically in Fig.~\ref{fig:fluctuation_diagrams} \textbf{(a)}.
	Finally, the last line yields the Gaussian-level diamagnetic diagram originally considered by Ref.~\cite{Boyack-Boyack-RestoringGaugeInvariance-2018}, which is shown graphically in  Fig.~\ref{fig:fluctuation_diagrams} \textbf{(e)}.
	In contrast to the AL diagram shown in Fig.\ref{fig:fluctuation_diagrams} \textbf{(d)}, all of these involve summation over a single flavor of fermions $\sfa$.
	
	\subsection{Bare electromagnetic vertices}
	
	The fluctuation corrections involve the bare electromagnetic vertices, $\gamma^{\mu}_{\sfa}$ and $\gamma^{\mu\nu}_{\sfa}$.
	These are found by reading off the paramagnetic and diamagnetic four-currents that appear in the free theory after minimal substitution \cite{Schrieffer-Schrieffer-TheorySuperconductivity-1999}.
	Performing one functional derivative yields the paramagnetic four-current with components
	\begin{equation}
			j^{0}_{\mathrm{p}}(x) = -e  \sum_{\sigma\sfa} \bar{\psi}_{\sigma\sfa}(x) \psi_{\sigma\sfa}(x),\quad \text{and} \quad
			j^{i}_{\mathrm{p}}(x) =  \iu \sum_{\sigma\sfa} \frac{e}{2m_{\sfa}} \bar{\psi}_{\sigma\sfa}(x) \left( \vec{\partial}_{i} - \, \backvec{\partial}_{i} \right) \psi_{\sigma\sfa}(x).
	\end{equation}
	Introducing the Fourier transforms of the fields yields
	\begin{equation}
		j^{\mu}_{\mathrm{p}}(q) =- e \sum_{k} \sum_{\sigma\sfa} \gamma^{\mu}_{\sfa}(k+q,k) \bar{\psi}_{\sigma\sfa}(k) \psi_{\sigma\sfa}(k+q),
	\end{equation}
	where the Fourier transform of the bare vertex defined in Eq.~\eqref{eq:bare_particle_and_hole_vertex} is
	\begin{equation}\label{eq:bare_vertex}
		\gamma^{\mu}_{\sfa}(k+q,k) \coloneqq \begin{cases}
			1 & \mu = 0 \\
			\displaystyle\frac{(\v{k} + \v{q}/2)^{i}}{m_{\sfa}} & \mu = i
		\end{cases}.
	\end{equation}
	The bare hole vertex is related to the bare particle vertex by $\tilde{\gamma}^{\mu}_{\sfa}(k+q,k) = - \gamma^{\mu}_{\sfa}(-k,-k-q)$ \cite{Boyack-Boyack-RestoringGaugeInvariance-2018}.
	Inspecting Eq.~\eqref{eq:bare_vertex} and Eq.~\eqref{eq:Free_Greens-function} it is clear that the \textit{bare} Ward-Takahashi identity is satisfied \cite{Schrieffer-Schrieffer-TheorySuperconductivity-1999}
	\begin{equation}
		q_{\mu} \gamma^{\mu}_{\sfa}(k+q,k) =  G_{0,\sfa}^{-1}(k+q) - G_{0,\sfa}^{-1}(k),
	\end{equation}
	where the contraction on the left-hand side is performed with the metric $\mathfrak{g}$.
	
	To relate the real-space vertex $\gamma^{\mu}_{\sfa}(z_1,z_2;x)$ in Eq.~\eqref{eq:bare_particle_and_hole_vertex} to its Fourier-components in Eq.~\eqref{eq:bare_vertex}, one can start from the definition of the vertex as a time-ordered expectation value of the paramagnetic current operator with the creation and annihilation operators of electrons \cite{Schrieffer-Schrieffer-TheorySuperconductivity-1999}
	\begin{align}
		\Lambda^{\mu}_{0}(x,y;z) &\coloneqq \sum_{\sigma \sfa} \langle T_{\tau} \left( j^{\mu}_{\mathrm{p}}(z) c_{\sigma\sfa}^{\mathstrut}(x) c^{\dagger}_{\sigma\sfa}(y) \right) \rangle \label{eq:vertex_operator_def_1} \\
		&\equiv -e \sum_{\sigma \sfa} \int \D x' \int \D y' G_{0,\sfa}(x,x') \gamma_{\sfa}^{\mu}(x',y';z) G_{0,\sfa}(y',y). \label{eq:vertex_operator_def_2}
	\end{align}
	Inserting the Fourier representation of the current in Eq.~\eqref{eq:vertex_operator_def_1} and using Wick's theorem permits comparing the result with Eq.~\eqref{eq:vertex_operator_def_2}.
    This yields
	\begin{equation}\label{eq:bare_vertex_Fourier}
		\gamma^{\mu}_{\sfa}(z_1,z_2;x) = \frac{1}{(\beta V)^2} \sum_{kq} \e^{\iu k \cdot (z_1-z_2) + \iu q \cdot ( z_1 - x)} \gamma^{\mu}_{\sfa}(k+q,k).
	\end{equation}
	
	The diamagnetic current vertex $\gamma^{\mu\nu}_{\sfa}(z_1,z_2;x,x')$ is found by
	performing two functional derivatives of the bare fermion Green's function with respect to $A$ according to its definition in Eq.~\eqref{eq:bare_diamagnetic_vertex}.
	This yields
	\begin{equation}\label{eq:gamma_munu}
		e^2\gamma^{\mu\nu}_{\sfa}(z_1,z_2;x,x') \equiv \frac{\delta^2 G_{0,\sfa}^{-1}[A](z_1,z_2)}{\delta A_{\mu}(x) \delta A_{\nu}(x')}  \biggr\lvert_{A=0} = - \delta(z_1-z_2) \delta(z_1-x) \delta(x-x') \frac{e^2}{m_{\sfa}} \delta^{\mu\nu} \left( 1 - \delta^{\mu0} \right).
	\end{equation} 
	Due to its local structure, it is most convenient to work with this vertex directly in real space.
	We now write out all the diagrams explicitly in real space, followed by their Fourier representations.
	
	\subsection{AL diagram}
	
	The AL diagram is found by computing the trace in Eq.~\eqref{eq:K_fluctuation_threept} with the vertex in Eq.~\eqref{eq:Lambda_mu}.
	This yields
	\begin{align}
		K^{\mu\nu}_{\mathrm{AL}}(x,x') &= - \sum_{abcd} \int \prod_{i=1}^{4} \D y_{i}  L_{ab}(y_1,y_2) \Lambda^{\mu}_{bc}(y_2,y_3;x) L_{cd}(y_3,y_4) \Lambda^{\nu}_{da}(y_4,y_1;x')  \notag \\
		&= - 4 e^2 \sum_{abcd} \sum_{\sfa\sfb} s_{b\sfa} s_{c\sfa} s_{d\sfb} s_{a\sfb} \int \prod_{i=1}^{4} \D y_{i} \int \prod_{j=1}^{4} \D z_{j} L_{ab}(y_1,y_2) L_{cd}(y_3,y_4) \notag \\
		&\hspace{20em}\times G_{0,\sfa}(y_2,z_1) \gamma^{\mu}_{\sfa}(z_1,z_2;x) G_{0,\sfa}(z_2,y_3) \tilde{G}_{0,\sfa}(y_3,y_2)  \\
		\phantom{\int\prod_{j=1}^{4}}&\hspace{20em}\times  G_{0,\sfb}(y_4,z_3) \gamma^{\mu}_{\sfb}(z_3,z_4;x) G_{0,\sfb}(z_4,y_1) \tilde{G}_{0,\sfb}(y_1,y_4). \notag 
	\end{align}
	Inserting the Fourier transforms of the propagators in Eq.~\eqref{eq:L_inv_explicit} and \eqref{eq:Fourier_transformed_G}, as well as the bare vertex in Eq.~\eqref{eq:bare_vertex_Fourier} yields
	\begin{equation}
		K^{\mu\nu}_{\mathrm{AL}}(x,x') = \frac{1}{\beta V } \sum_{q} \e^{\iu q \cdot(x-x')} K^{\mu\nu}_{\mathrm{AL}}(q).
	\end{equation}
	The expression for $K^{\mu\nu}_{\mathrm{AL}}(q)$ is simplified by performing the integrals over positions, which ultimately yields
	\begin{equation}
		\begin{split}
			K^{\mu\nu}_{\mathrm{AL}}(q) &= - \frac{4e^2}{(\beta V)^3} \sum_{k k' p} \sum_{\sfa\sfb} \left(S^{\mathsf{T}} L(p-q) S \right)_{\sfa\sfb} \left( S^{\mathsf{T}} L(p) S \right)_{\sfb\sfa} G_{0,\sfa}(k-q) \gamma^{\mu}_{\sfa}(k-q,k) G_{0,\sfa}(k)  G_{0,\sfa}(p-k) \\
			&\hspace{20.5em}\times G_{0,\sfb}(k') \gamma^{\nu}_{\sfb}(k',k'-q) G_{0,\sfb}(k'-q) G_{0,\sfb}(p-k'). \phantom{\sum_{k}}
		\end{split}
	\end{equation}
    
	\subsection{MT diagram}
	
	The MT diagram is found by computing the trace in Eq.~\eqref{eq:K_fluctuation_fourpt} with the part of $\Gamma^{\mu\nu}$ appearing in the third line of Eq.~\eqref{eq:Gamma_munu}.
	This yields
	\begin{align}
		K^{\mu\nu}_{\mathrm{MT}}(x,x') &= \sum_{ab} \int \prod_{i=1}^{2} \D y_{i} L_{ab}(y_1,y_2) \left[\Gamma_{\mathrm{MT}}\right]_{ba}^{\mu\nu}(y_2,y_1;x,x') \notag \\
		\begin{split}
			&= - 2e^2 \sum_{ab} \sum_{\sfa} s_{a\sfa} s_{b\sfa} \int \prod_{i=1}^{2} \D y_{i} \int \prod_{j=1}^{4} \D z_{i} L_{ab}(y_1,y_2) \\
			&\hspace{10em}\times G_{0,\sfa}(y_2,z_1) \gamma^{\nu}_{\sfa}(z_1,z_2;x') G_{0,\sfa}(z_2,y_1) \tilde{G}_{0,\sfa}(y_1,z_3) \tilde{\gamma}^{\mu}_{\sfa}(z_3,z_4;x) \tilde{G}_{0,\sfa}(z_4,y_2).
		\end{split}
	\end{align}
	Inserting the Fourier representations and performing the real space integrals results in
	\begin{equation}
		K^{\mu\nu}_{\mathrm{MT}}(x,x') = \frac{1}{\beta V } \sum_{q} \e^{\iu q \cdot(x-x')} K^{\mu\nu}_{\mathrm{MT}}(q),
	\end{equation}
	with
	\begin{equation}
		\begin{split}
			K^{\mu\nu}_{\mathrm{MT}}(q) &= \frac{2e^2 }{(\beta V)^2} \sum_{k p}  \sum_{\sfa} \left( S^{\mathsf{T}} L(p) S \right)_{\sfa\sfa}  G_{0,\sfa}(k-q) \gamma_{\sfa}^{\mu}(k-q,k)  G_{0,\sfa}(k) \\
			&\hspace{12.1em} \times G_{0,\sfa}(p-k) \gamma_{\sfa}^{\nu}(p+q-k,p-k) G_{0,\sfa}(p+q-k).
		\end{split}
	\end{equation}
		
	\subsection{DOS diagram}
	
	The DOS diagram is found by computing the trace in Eq.~\eqref{eq:K_fluctuation_fourpt} with the part of $\Gamma^{\mu\nu}$ appearing in the first and second line of Eq.~\eqref{eq:Gamma_munu}.
	This yields
	\begin{align}
		K^{\mu\nu}_{\mathrm{DOS}}(x,x') &= \sum_{ab} \int \prod_{i=1}^{2} \D y_{i} L_{ab}(y_1,y_2) \left[\Gamma_{\mathrm{DOS}}\right]_{ba}^{\mu\nu}(y_2,y_1;x,x') \notag \\
		\begin{split}
			&= - 2e^2 \sum_{ab} \sum_{\sfa} s_{a\sfa} s_{b\sfa} \int \prod_{i=1}^{2} \D y_{i} \int \prod_{j=1}^{4} \D z_{i} L_{ab}(y_1,y_2) \\
			&\hspace{8em}\times \bigg[ G_{0,\sfa}(y_2,z_1) \gamma^{\mu}_{\sfa}(z_1,z_2;x) G_{0,\sfa}(z_2,z_3) \gamma^{\nu}_{\sfa}(z_3,z_4;x') G_{0,\sfa}(z_4,y_1) \tilde{G}_{0,\sfa}(y_1,y_2) \\
			&\hspace{8.5em}+G_{0,\sfa}(y_2,z_1) \gamma^{\nu}_{\sfa}(z_1,z_2;x') G_{0,\sfa}(z_2,z_3) \gamma^{\mu}_{\sfa}(z_3,z_4;x) G_{0,\sfa}(z_4,y_1) \tilde{G}_{0,\sfa}(y_1,y_2) \bigg].
		\end{split}
	\end{align}
	Inserting the Fourier representations yields
	\begin{equation}
		K^{\mu\nu}_{\mathrm{DOS}}(x,x') = \frac{1}{\beta V } \sum_{q} \e^{\iu q \cdot(x-x')} K^{\mu\nu}_{\mathrm{DOS}}(q),
	\end{equation}
	with
	\begin{equation}
		\begin{split}
			K^{\mu\nu}_{\mathrm{DOS}}(q)&=  \frac{2 e^2}{(\beta V)^2} \sum_{k p}  \sum_{\sfa} \left( S^{\mathsf{T}} L(p) S \right)_{\sfa\sfa} G_{0,\sfa}(k-q) \gamma^{\mu}_{\sfa}(k-q,k) G_{0,\sfa}(k)  G_{0,\sfa}(p-k) G_{0,\sfa}(k) \gamma^{\nu}_{\sfa}(k,k-q) \\
			&\,+  \frac{2 e^2}{(\beta V)^2} \sum_{k p} \sum_{\sfa} \left( S^{\mathsf{T}} L(p) S \right)_{\sfa\sfa} G_{0,\sfa} (k) \gamma^{\mu}_{\sfa}(k,k+q) G_{0,\sfa}(k+q) \gamma_{\sfa}^{\nu}(k+q,k) G_{0,\sfa}(k) G_{0,\sfa}(p-k) 
		\end{split}
	\end{equation}
	
	\subsection{DIA diagram}
	
	Finally, the DIA diagram is found by computing the trace in Eq.~\eqref{eq:K_fluctuation_fourpt} with the part of $\Gamma^{\mu\nu}$ appearing in the last line of Eq.~\eqref{eq:Gamma_munu}.
	This yields
	\begin{align}
			K^{\mu\nu}_{\mathrm{DIA}}(x,x') &= \sum_{ab} \int \prod_{i=1}^{2} \D y_{i} L_{ab}(y_1,y_2) \left[\Gamma_{\mathrm{DIA}}\right]_{ba}^{\mu\nu}(y_2,y_1;x,x') \notag \\ 
            \begin{split}
                &= 2 e^2 \sum_{ab} \sum_{\sfa} s_{a\sfa} s_{b\sfa} \int \prod_{i=1}^{2} \D y_{i} \int \prod_{j=1}^{2} \D z_{j} L_{ab}(y_1,y_2) \\
    			&\hspace{10em} \times G_{0,\sfa}(y_2,z_1) \gamma_{\sfa}^{\mu\nu}(z_1,z_2;x,x') G_{0,\sfa}(z_2,y_1) \tilde{G}_{0,\sfa}(y_1,y_2).
		      \end{split}
	\end{align}
	Inserting the Fourier transform of the propagators together with Eq.~\eqref{eq:gamma_munu} now directly yields $K^{\mu\nu}(x,x') =  \delta(x-x') K^{\mu\nu}_{\mathrm{DIA}}(q)$ \cite{Boyack-Boyack-RestoringGaugeInvariance-2018}, that is 
	\begin{equation}
		K^{\mu\nu}_{\mathrm{DIA}}(x,x') = \delta(x-x') \frac{2e^2}{(\beta V)^2} \sum_{k p }  \sum_{\sfa} \frac{1}{m_{\sfa}} \delta^{\mu\nu}(1 - \delta^{\mu 0}) \left( S^{\mathsf{T}} L(p) S \right)_{\sfa\sfa} \left[ G_{0,\sfa}(k) \right]^2 G_{0,\sfa}(p-k).
	\end{equation}
    
	\twocolumngrid
	
	\bibliography{references_fluctcon}
    
\end{document}